\documentclass[twocolumn,aps,amsmath,amssymb,amsfonts,groupedaddress,pra]{revtex4-1} 
\usepackage{graphicx}% Include figure files
\usepackage{dcolumn}% Align table columns on decimal point
\usepackage{bm}% bold math

\usepackage[utf8]{inputenc} % set input encoding (not needed with XeLaTeX) 
\usepackage{float}

\usepackage{array} % for better arrays (eg matrices) in maths 
\usepackage{verbatim} % adds environment for commenting out blocks of text & for better verbatim 
\usepackage[unicode=true,pdfusetitle, 
 bookmarks=true,bookmarksnumbered=false,bookmarksopen=false,
 breaklinks=false,pdfborder={0 0 1},backref=false,colorlinks=false]
 {hyperref}
\usepackage[usenames,dvipsnames]{xcolor}
\hypersetup{colorlinks=true, linkcolor=Maroon, citecolor=OliveGreen, filecolor=magenta, urlcolor=Blue} 
\newcommand{\Liquid}{LIQ$Ui\ket{}$\ }
\newcommand{\LiquidB}{LIQ$Ui\ket{}$}

\usepackage{xypic}
\xyoption{matrix}
\xyoption{frame}
\xyoption{arrow}
\xyoption{arc}

\usepackage{ifpdf}
\ifpdf
\else
\PackageWarningNoLine{Qcircuit}{Qcircuit is loading in Postscript mode.  The Xy-pic options ps and
dvips will be loaded.  If you wish to use other Postscript drivers for Xy-pic, you must modify the
code in Qcircuit.tex}
\xyoption{ps}
\xyoption{dvips}
\fi

\entrymodifiers={!C\entrybox}

\newcommand{\ket}[1]{{\left\vert{#1}\right\rangle}}
\newcommand{\qw}[1][-1]{\ar @{-} [0,#1]}
\newcommand{\qwx}[1][-1]{\ar @{-} [#1,0]}
\newcommand{\gate}[1]{*+<.6em>{#1} \POS ="i","i"+UR;"i"+UL **\dir{-};"i"+DL **\dir{-};"i"+DR **\dir{-};"i"+UR **\dir{-},"i" \qw}
\newcommand{\meter}{*=<1.8em,1.4em>{\xy ="j","j"-<.778em,.322em>;{"j"+<.778em,-.322em> \ellipse ur,_{}},"j"-<0em,.4em>;p+<.5em,.9em> **\dir{-},"j"+<2.2em,2.2em>*{},"j"-<2.2em,2.2em>*{} \endxy} \POS ="i","i"+UR;"i"+UL **\dir{-};"i"+DL **\dir{-};"i"+DR **\dir{-};"i"+UR **\dir{-},"i" \qw}
\newcommand{\control}{*!<0em,.025em>-=-<.2em>{\bullet}}
\newcommand{\ctrl}[1]{\control \qwx[#1] \qw}
\newcommand{\targ}{*+<.02em,.02em>{\xy ="i","i"-<.39em,0em>;"i"+<.39em,0em> **\dir{-}, "i"-<0em,.39em>;"i"+<0em,.39em> **\dir{-},"i"*\xycircle<.4em>{} \endxy} \qw}

\newcommand{\rstick}[1]{*!L!<-.5em,0em>=<0em>{#1}}
\newcommand{\lstick}[1]{*!R!<.5em,0em>=<0em>{#1}}
\newcommand{\Qcircuit}{\xymatrix @*=<0em>}

\begin{document}
\title{Low-distance Surface Codes under Realistic Quantum Noise}
\author{Yu Tomita$^\dagger$}
\author{Krysta M. Svore$^*$}
\affiliation{$^\dagger$School of Computational Science and Engineering, Georgia Institute of Technology, Atlanta, GA (USA)\\
$^*$Quantum Architectures and Computation Group, Microsoft Research, Redmond, WA (USA)}

\begin{abstract}

We study the performance of distance-three surface code layouts under realistic multi-parameter noise models.
We first calculate their thresholds under depolarizing noise.
We then compare a Pauli-twirl approximation of amplitude and phase damping to amplitude and phase damping.
We find the approximate channel results in a pessimistic estimate of the logical error rate, indicating the realistic threshold may be higher than previously estimated. 
%Finally, we analyze the layouts under uniform magnetic field noise.
From Monte-Carlo simulations, we identify experimental parameters for which these layouts admit reliable computation.
Due to its low resource cost and superior performance, we conclude that the 17-qubit layout should be targeted in early experimental implementations of the surface code.
We find that architectures with gate times in the 5--40 ns range and $T_1$ times of at least 1--2 $\mu$s will exhibit improved logical error rates with a 17-qubit surface code encoding.

\end{abstract}
\maketitle

\section{Introduction}

Topological quantum error-correcting codes are a leading approach to scalable fault-tolerant quantum computation \cite{Fowler092012,Lidar2013}. 
The most practical topological code to date is the surface code, which calls for a 2-D planar qubit layout with only nearest-neighbor interactions \cite{BravyiKitaev98,Kitaev2003,RaussendorfPRL2007,FowlerPRA2009}.
It has been shown to allow error rates up to a threshold of approximately 1\%~\cite{Fowler052012,Fowler092012,Wang2011,Stephens2013}. 
Several quantum architectures, including superconducting devices \cite{Helmer2009,Divincenzo2009} and ion traps \cite{KMW,monroe2014,TrueNJP2011,wright2012}, are suitable for realizing the surface code. 
Recent experiments on superconducting qubits have even demonstrated error rates in the required range \cite{Martinis14}.
%,DiCarlo_email}.

Until recently, the threshold for the surface code has been primarily calculated for the depolarizing channel \cite{Fowler052012, Fowler092012, Wang2011, Stephens2013}.  
Simulation of the surface code and the depolarizing channel requires only Clifford operations and Pauli measurements on stabilizer states, allowing efficient simulation on a classical computer under the Gottesman-Knill theorem~\cite{Gottesman1998,Aaronson2004}.

It has been shown that realistic quantum noise such as decoherence can be sufficiently approximated by a depolarizing noise model parametrized by a method such as Pauli twirling \cite{Silva2008, Nielsen_book}, enabling efficient simulation.
Simulations of the surface code with noise based on Pauli-twirl approximations have been performed for several superconductor architectures \cite{Ghosh2012}.
Other studies have achieved efficient classical simulation of realistic noise models by using Clifford gates to approximate arbitrary gates \cite{Cory012013, Cory2013} and amplitude damping \cite{Mauricio2013}. 
%The former model has been compared with a model of arbitrary quantum errors on a circuit segment of the surface code~\cite{Cory2013} KMS:AND FOUND TO BE WHAT?.

More recently, it has been shown that the surface code threshold is significantly degraded in the presence of qubit leakage in conjunction with depolarizing noise~\cite{Fowler082013}.
It has also been shown to achieve arbitrary reliability given modest additional qubit resources under local many-qubit errors and non-local two-qubit errors \cite{Fowler14}.
A recent study has determined a threshold for the surface code considering correlated errors and the coupling between qubits and the environment by formulating the problem as an Ising model \cite{Jouzdani14}.

In all cases, the thresholds have been calculated for a standard surface code layout.
Variations of the surface code layout have been proposed \cite{Bombin07,Horsman2012} that reduce the qubit and gate resources necessary for implementation.
To the best of our knowledge, the thresholds for these modified surface code layouts have not been analyzed.
In addition, studies of the threshold under realistic (non-Clifford) noise models have been limited due to the exponential cost of simulation. 
With device error rates rapidly approaching the surface code threshold, it is timely to investigate the performance and requirements of low-distance surface code layouts for near-term experimental implementation.

In this work, we determine the threshold for distance-three surface code layouts under depolarizing and realistic noise models.
We study the layouts under an amplitude and phase damping channel and an approximation of the channel using Pauli twirling \cite{Ghosh2012}.
%, and uniform magnetic field.
Our studies demand simulation of non-Clifford operations, which requires memory exponential in the number of qubits.
We use the \Liquid \cite{Liquid_Documentation} software architecture for our simulations.
% we are able to simulate 
%up to 31 qubits with only 32GB RAM, enabling simulation of 
%realistic noise models on distance-three surface code layouts.
% for the first time.
We also outline parameter regimes that enable reliable quantum error correction for low-distance surface codes and present a decoder based on a small lookup table optimized for distance-three layouts and limited classical computation.

Our paper is organized as follows.  Section \ref{sec:codes} briefly reviews the surface code and three layouts for the distance-three code.  We introduce our decoding method, based on a small lookup table, in Section \ref{sec_decoding}.  Section \ref{sec:noise} describes the realistic noise models and their approximations.
Our experimental methodology is introduced in Section \ref{sec_steps}.
In Section \ref{sec:results}, we present our surface code simulation results.
Finally, we conclude in Section \ref{sec:conclude}.

%% -------------------------------------------------------------------------------------------------

\section{Low-distance Surface codes}\label{sec:codes}

%% -------------------------------------------------------------------------------------------------

The surface code is a stabilizer code arranged on a 2-D lattice with nearest-neighbor interactions \cite{BravyiKitaev98}.
It encodes a single logical qubit in a number of physical qubits that is determined by the code distance $d$ and desired layout (described below).
Through repeated measurement of its stabilizer generators, the surface code in conjunction with a classical decoding algorithm can detect errors and subsequently correct up to $\lfloor (d-1)/2 \rfloor$ physical errors.
The distance $d$ dictates the length of the shortest undetectable error chain and in turn is also the length of the shortest logical operator.
For an excellent review of the surface code, we refer the reader to \cite{Fowler092012}.

\subsection{25-qubit Layout}

\begin{figure}[bt] 
	\begin{tabular}{ccc}
	\includegraphics[width=0.16\textwidth]{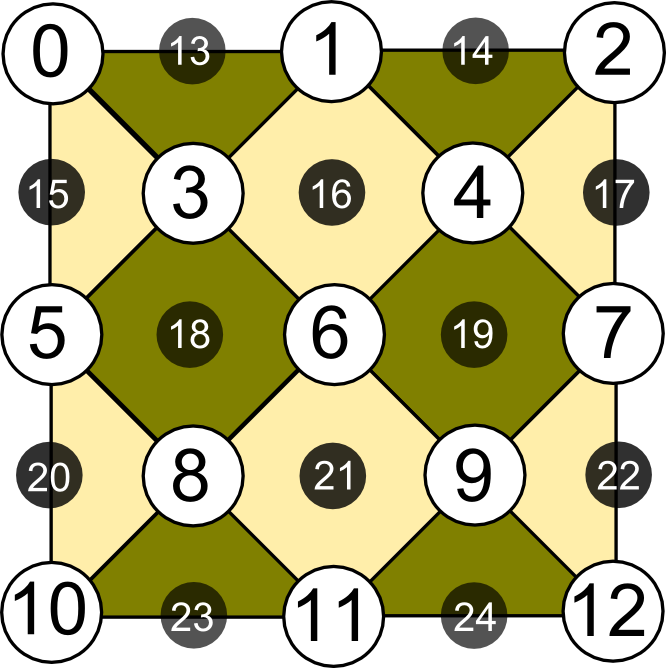} &
	\includegraphics[width=0.16\textwidth]{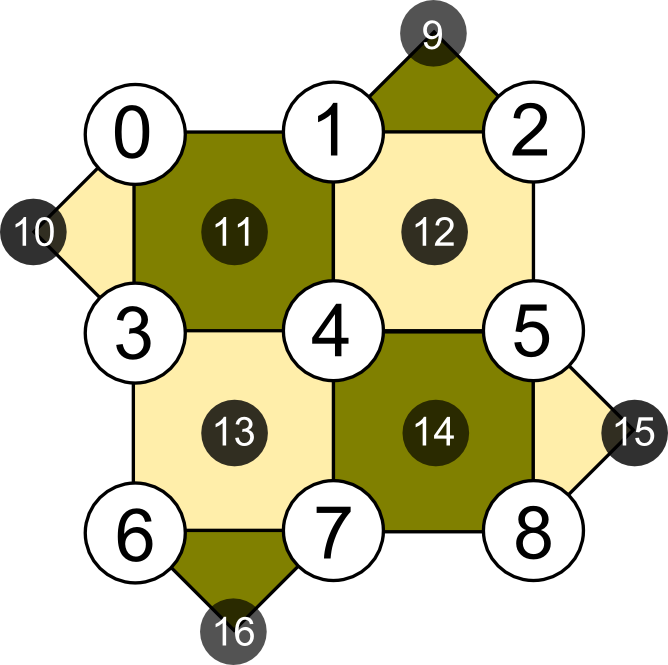} &
	\includegraphics[width=0.13\textwidth]{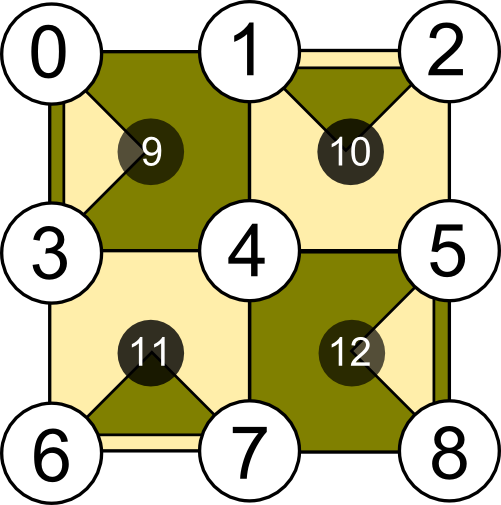}\\
(a) Surface-25 & (b) Surface-17 & (c) Surface-13 \\
	\end{tabular}
	\caption{Distance-three surface code layouts with (a) 25, (b) 17, and (c) 13 qubits. White circles represent data qubits; black circles represent syndrome qubits.  Dark square and triangle patches represent $X$ stabilizers; light patches represent $Z$ stabilizers. The layered patches on Surface-13 indicate use of the syndrome qubit first to measure a four-qubit stabilizer and then to measure a two-qubit stabilizer.}\label{fig_surfacecodes}
\end{figure}

We study three different distance $d=3$ layouts, shown in Figure \ref{fig_surfacecodes}.
We begin by discussing the standard layout, referred to as Surface-25, shown in Figure \ref{fig_surfacecodes}(a).
It uses a $(2d-1)\times(2d-1)$ square grid of qubits with a smooth and rough boundary~\cite{Kitaev2003}. 
For $d=3$, the grid contains $25$ qubits of which $13$ \textit{data} qubits (large white circles) are used to encode the logical qubit and $12$ \textit{syndrome} qubits (small black circles) are used to extract the error syndromes by way of stabilizer measurements.

Surface-25 is simultaneously stabilized by the group of stabilizer generators listed in Table \ref{tab_stab}. 
In Fig.~\ref{fig_surfacecodes}, the $Z$ stabilizers are represented by light (yellow) patches and the $X$ stabilizers are represented by dark (green) patches, where each patch represents a tensor product of $Z$ (or $X$) operators on the data qubits surrounding the patch.
%For example, one of the $X$ stabilizers is given by $X_1X_3X_4X_6$.

A logical $X$ operator $X_L$ is defined as a chain of physical $X$ operations between two data qubits on opposite smooth boundaries (top and bottom edges).
The chain is allowed to cross any $Z$ stabilizer patch and follow any edge of an $X$ stabilizer patch.
A logical $Z$ operator $Z_L$ is defined analogously as a chain of physical $Z$ operations between two data qubits on opposite rough boundaries (left and right edges). 
Table \ref{tab_stab} lists one possible logical $X$ and $Z$ operator. 
There are $2^G$ equivalent logical operators for each logical Pauli operator ($X$ and $Z$), where $G$ is the number of stabilizer generators for the given surface code. 
Since $X_L$ and $Z_L$ commute with all of the stabilizers and cannot be written as a product of them, logical errors, which come in the form of logical operators, cannot be detected by the code. 

\begin{figure}[tb!] 	
	\includegraphics[width=0.4\textwidth]{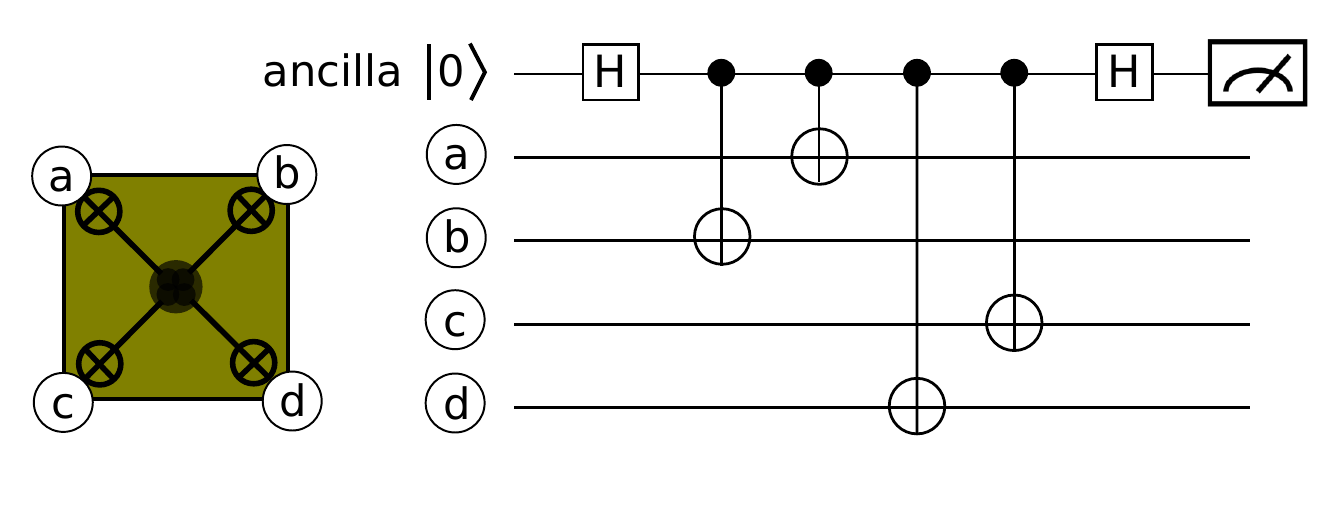} \\ 
(a)\\

	\includegraphics[width=0.4\textwidth]{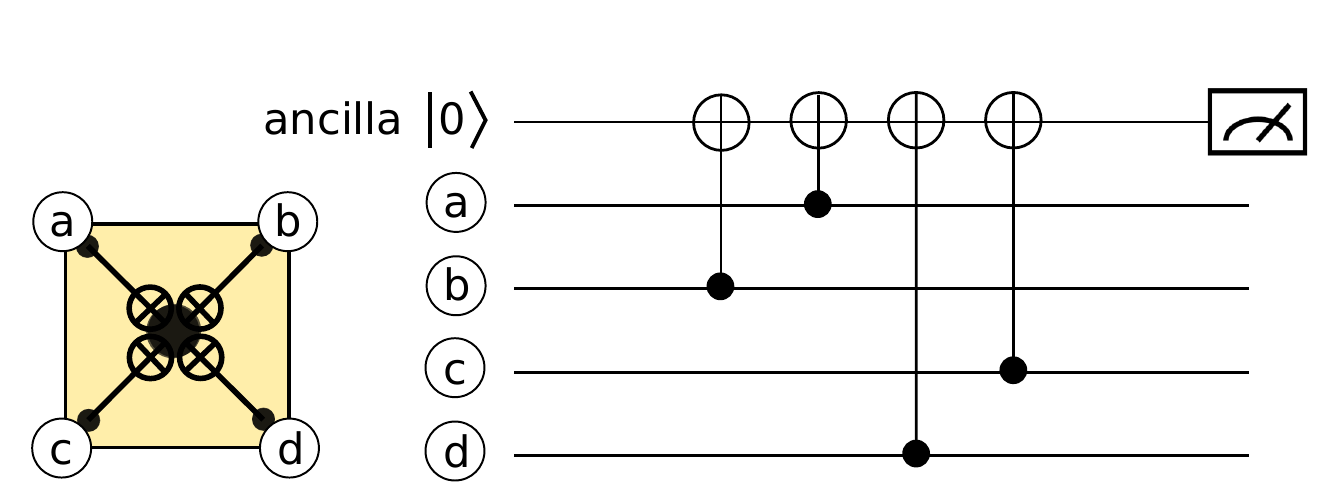} \\
(b)	
	\caption{Standard quantum circuits to measure stabilizers (a) $X_aX_bX_cX_d$ and (b) $Z_aZ_bZ_cZ_d$.}\label{fig_oldorder}
\end{figure}

The surface code detects errors through the eigenvalues of the stabilizers.
A bit-flip (phase-flip) on a data qubit will change the eigenvalue of adjacent $Z$ ($X$) stabilizers. 
To extract an eigenvalue, also referred to as an \textit{error syndrome}, a given stabilizer is measured.
Figure \ref{fig_oldorder} shows the standard quantum circuit for measuring the stabilizers~\cite{Fowler092012,Stephens2013},
where data qubit $b$ corresponds to the top (north) qubit and $c$ corresponds to the bottom (south) qubit of each diamond patch in Fig.~\ref{fig_surfacecodes}(a).

The circuit begins with CNOT gates that propagate error information from the data qubits $a$,$b$,$c$,$d$ to the syndrome qubit (black circle).
CNOT gates are performed in the order: top ($b$); left ($a$); right ($d$); bottom ($c$).
Cyclic orders, such as a clockwise or counter-clockwise, i.e., $bdca$, fail to maintain commutation of nearby stabilizers, which in turn can cause random measurement outcomes \cite{Fowler092012}.  
Thus the order of CNOT gates is required to follow an ``S" or ``Z" shape. 

The syndrome qubit is then measured to extract the eigenvalue of the stabilizer. 
These error syndromes are input to a classical decoding algorithm to determine an appropriate correction operator.
Details of our decoding algorithm are given in Section \ref{sec_decoding}.
The total number of operations in a given round of stabilizer measurements for the surface code is given in Table \ref{tab_counts}.

\begin{table}[tb]
	\caption{Number of operations in one round of the surface code for distance-three layouts. 
%KMS: UPDATE VALUES IN TABLE. YT: UPDATED-- I GATES INCLUDE ALL FOUR LATENCIES OF I (LOCAL, CX, RESET, MEAS)
}
\label{tab_counts}
\begin{tabular}{| c | c | c | c | c | c | c |}
\hline
Code		& CNOT & $I$ & $H$ & Meas. & Prep. & Depth \\ \hline\hline
Surface-13 	& $24$ & $99$ & $8$ & $8$ & $8$ & $14$ \\ \hline
Surface-17 	& $24$ & $56$ & $8$ & $8$ & $8$ & $8$ \\  \hline
Surface-25 	& $40$ & $72$ & $12$ & $12$ & $12$ & $8$ \\ \hline 
\end{tabular}
\end{table}

\begin{table}[tb!]
\caption{List of $X$ and $Z$ stabilizers and logical $X_L$ and $Z_L$ operators for Surface-13, 17, and 25.}\label{tab_stab}
\begin{tabular}{| c | c | c | c |}
  \hline
  \multicolumn{2}{|c|}{Surface-25}		 		& \multicolumn{2}{c|}{Surface-13, Surface-17} \\
  \hline
  \hline
  $X$ Stabilizers 		& $Z$ Stabilizers 		& $X$ Stabilizers 		& $Z$ Stabilizers \\ \hline 
  $X_0 X_1 X_3$ 		& $Z_0 Z_3 Z_5$ 		& $X_0 X_1 X_3 X_4$ 	& $Z_0 Z_3$ \\  
  $X_1 X_2 X_4$ 		& $Z_1 Z_3 Z_4 Z_6$ 	& $X_1 X_2 $ 			& $Z_1 Z_2 Z_4 Z_5$ \\   
  $X_3 X_5 X_6 X_8 $ 	& $Z_2 Z_4 Z_7$ 		& $X_4 X_5 X_7 X_8 $ 	& $Z_3 Z_4 Z_6 Z_7$ \\  
  $X_4 X_6 X_7 X_9$ 	& $Z_5 Z_8 Z_{10}$  	& $X_6 X_7$ 			& $Z_5 Z_8$ \\  
  $X_8 X_{10} X_{11}$ 	& $Z_6 Z_8 Z_9 Z_{11}$	& & \\ 
  $X_9 X_{11} X_{12}$	& $Z_7 Z_9 Z_{12}$		& & \\ \hline
  \hline
  Logical $X$				& Logical $Z$ 			& Logical $X$ 			& Logical $Z$ \\ \hline
  $X_0 X_5 X_{10}$ 		& $Z_0 Z_1 Z_2$  		& $X_2 X_4 X_6$			& $Z_0 Z_4 Z_8$ \\ \hline 
\end{tabular}
\end{table}

\subsection{13- and 17-qubit Layouts}

\begin{figure}[tb!] 
	\includegraphics[width=0.5\textwidth]{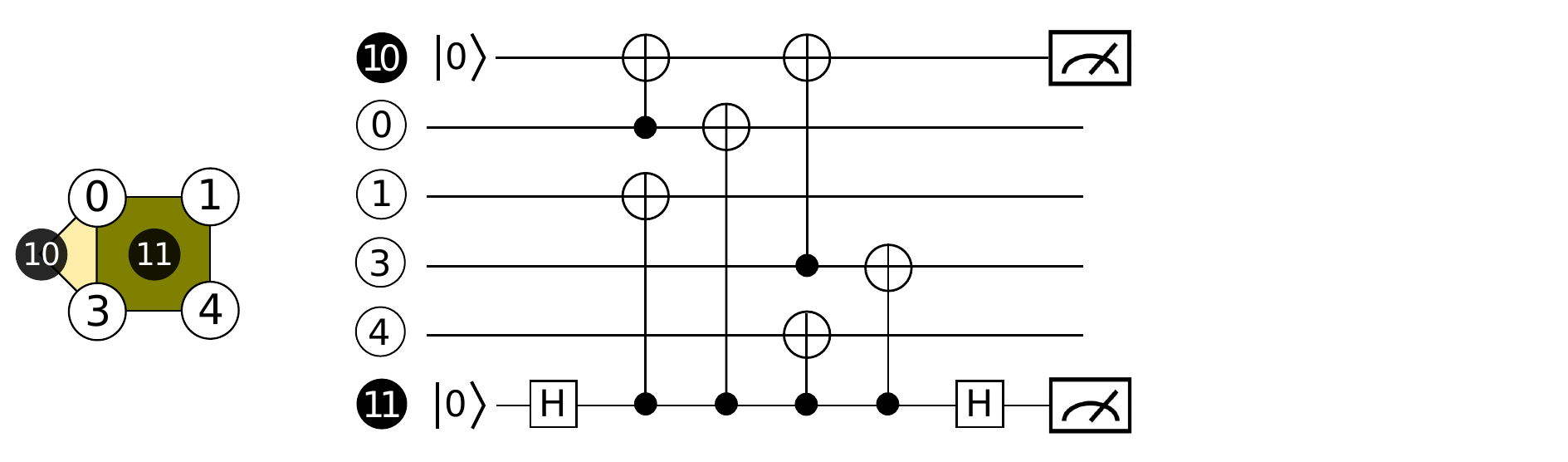} \\ 
	(a)\\

	\includegraphics[width=0.5\textwidth]{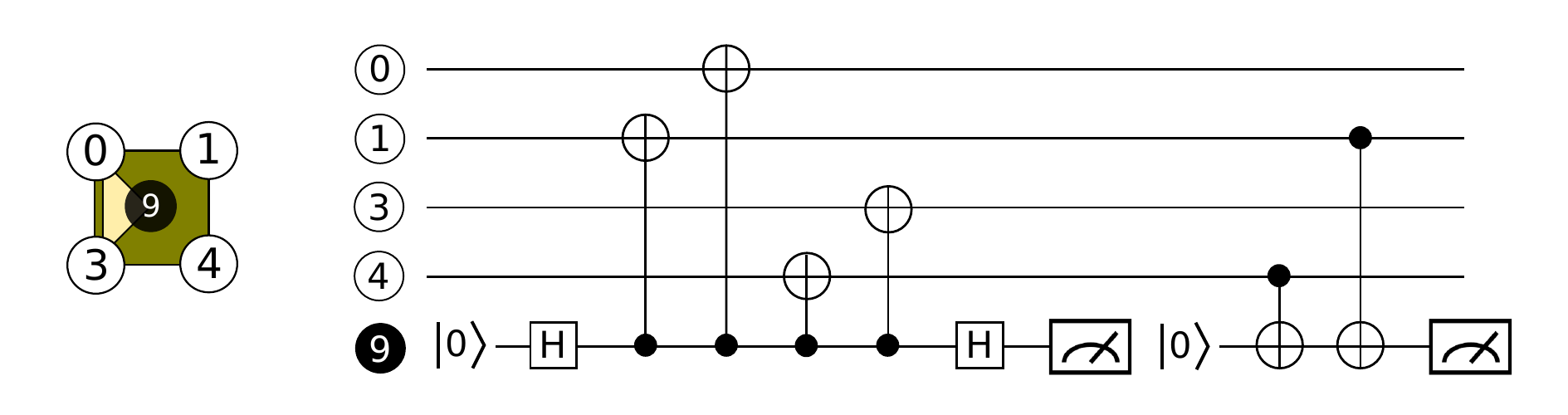} 	\\
(b)
	\caption{Quantum circuits for measuring $X_0X_1X_3X_4$ and $Z_0Z_3$ in (a) Surface-17 and (b) Surface-13.}\label{fig_17_13_circ}
\end{figure}

The number of qubits in Surface-25 can be reduced while maintaining the same code distance by rotating it clockwise by $45$ degrees and removing the four corner data qubits \cite{Bombin07,Horsman2012}, shown in Fig.~\ref{fig_surfacecodes}(b).  
The number of data qubits is reduced from $13$ to $9$ and the number of syndrome qubits is reduced to $8$ for a total of $17$ qubits.  We call this layout Surface-17.  
The stabilizer generators contain weight-4 and weight-2 stabilizers (Table \ref{tab_stab}).
Figure \ref{fig_17_13_circ}(a) shows the circuit for a simultaneous weight-4 $X$ and weight-2 $Z$ stabilizer measurement.

A further reduction in qubits can be obtained by reusing the syndrome qubits \cite{Horsman2012}.
Surface-13 uses only $4$ syndrome qubits as shown in Fig.~\ref{fig_surfacecodes}(c).
Each syndrome qubit is used twice, once for $X$ stabilizer measurement and once for $Z$ stabilizer measurement.
Figure \ref{fig_17_13_circ}(b) contains the corresponding circuit for measuring a weight-4 $X$ stabilizer followed by a weight-2 $Z$ stabilizer.
Surface-13 reduces the number of qubits but increases the depth of a round by 4 time steps.
%KMS: ADD DIAGRAM TO SHOW DIFFERENCE FOR LAST ROUND? YT: FIGURE \ref{fig_17_13_circ} ADDED 
The depth and number of operations required for one round of the surface code for Surface-17 and 13 are given in Table \ref{tab_counts}.
The stabilizers and logical operations for these two layouts are listed in Table \ref{tab_stab}.

%In the case of Surface-17, $X_0 X_3 X_6$ and $X_0 X_3 X_7$ are both valid logical $X$ operators while $X_0 X_4 X_7$ is not. 
Despite having fewer stabilizers, Surface-17 and Surface-13 still remain distance-three surface codes~\cite{Bombin07,Horsman2012}.
Due to their reduction in resources by 32--48\%, these layouts are promising candidates for early experimental implementation.
In Section \ref{sec:results}, we determine which layout is most promising based on its threshold and resource costs.

\section{Decoding Method}\label{sec_decoding}

A standard method for mapping error syndromes to the most probable error chain is the minimum weight perfect matching algorithm \cite{Fowler052012, Edmonds65,Edmonds65b}.
It requires time $O(n)$ for $n$ detection events if executed serially, and $O(1)$ time if executed in parallel \cite{Fowler072013}.
The algorithm independently corrects $X$ and $Z$ errors by identifying the most likely error chain for each type such that the total chain weight is minimal.
The algorithm has recently been extended to handle correlations between $X$ and $Z$ errors, in which case the chains are not constructed independently \cite{Fowler13100863}.
Corrections are applied along the chain(s).
If after correction a chain of errors connecting two smooth (rough) boundaries remains, then a logical error has occurred.
If errors are assumed to be independent, then long chains will be exponentially unlikely.

\subsection{Lookup Table Decoder}

In this work we target first-generation implementations of a single qubit protected by a small surface code.
While the classical time and space requirements of the minimum weight perfect matching algorithm are modest, we further reduce the classical computational overhead by designing a lookup table based on the algorithm that can be implemented on a small classical device.
Our lookup table is designed to find the most probable low-weight error chain from a short history of error syndromes. 

%Assume the surface code undergoes noise such that errors are independent.
Consider the set of error syndromes that indicate an error after one full (noisy) round of the surface code, that is, those indicating a $-1$ eigenvalue.
Based on the error syndrome locations, the decoder determines the probable data-qubit error locations.
For example, consider a $Z$ error on qubit $4$ in Surface-17 (Fig.~\ref{fig_surfacecodes}(b)).
Given that no other errors occur, after one round of the surface code syndrome qubits $11$ and $14$ will indicate an error.
The decoder will determine the shortest error chain connecting these two syndromes includes data qubit $4$.  
To correct the error chain, $Z_4$ will be applied. 
%KMS: DO WE WANT A VISUAL FOR THIS AND THE NEXT EXAMPLE?

As another example, consider an $X$ error on qubit $6$. It will cause syndrome $13$ to indicate an error.
Since syndromes $10$ and $12$ do not indicate errors, the decoder will infer an error on either data qubit $6$ or $7$.  
In this case, the decoder can correct either $X_6$ or $X_7$ since $X_6X_7$ is a stabilizer.

An error syndrome may also occur due to a measurement error.
%If an error occurs during measurement, an error syndrome can due to an error on a syndrome qubit.
However, the decoder may interpret it as a data-qubit error.  
For example, consider a measurement error on qubit $11$.  The decoder will either apply $Z_0$ or $Z_3$ to ``correct" the error, thereby adding an error to a clean data qubit.

To improve identification of actual data-qubit errors, inference is performed based on several rounds of stabilizer measurements \cite{Fowler052012}.
Consider performing $r$ rounds of the surface code consecutively.
Instead of storing the syndromes for each round, we store the locations in time and space of the syndromes whose values change, or ``flip",  between the current and previous round.

For $r$ rounds, this requires storing a 3-D space-time array containing at most $s\times r$ values, where $s$ is the maximum number of syndrome changes in a round.
We refer to this 3-D array as the \textit{syndrome volume}, where dimension $r$ represents time.
The goal is to determine a correction operator (a product of $X$ and/or $Z$ operators) based on the syndrome volume such that the number of errors remaining after correction is minimized, in turn reducing the chance of forming a logical error chain.

\begin{figure}[tb] 
	\includegraphics[width=0.3\textwidth]{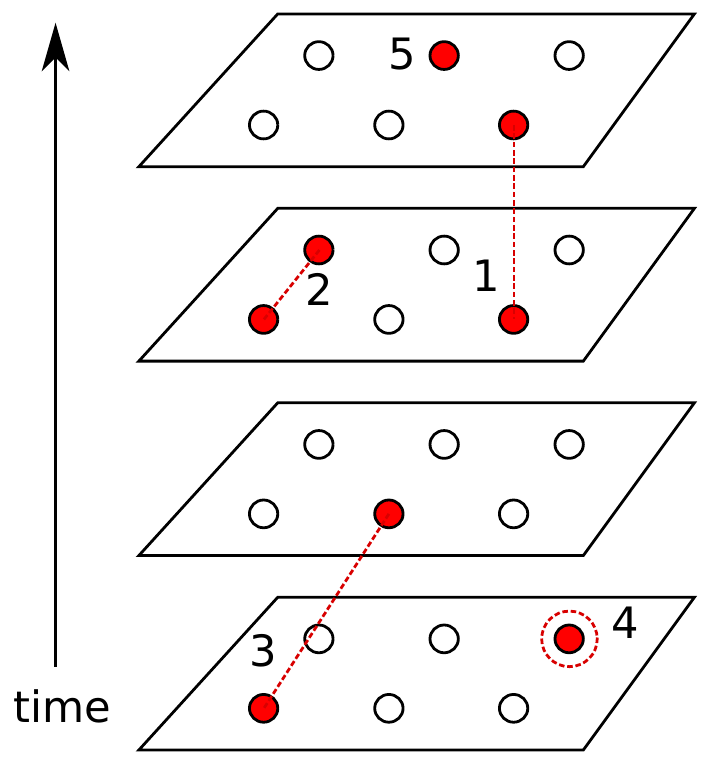} \\ 
	\caption{Lookup table decoding rules. Each circle represents a syndrome measurement. Red circles indicate ``flips". Five types of flips are shown: (1) measurement error, (2,3) paired flips indicating single-qubit error on data qubit, (4) single flip indicating one data-qubit error, and (5) undetermined flip. These numbers correspond to the rules in Section \ref{sec_decoding}.}\label{fig_decoding}
\end{figure}

Our lookup table is based on the fact that short error chains are more likely than long chains.
Assuming a syndrome volume contains $r$ rounds, we construct a lookup table based on the following rules (Figure \ref{fig_decoding} shows the rules visually): 
\begin{enumerate}
\itemsep 1pt
\parskip 0pt
	\item If the same syndrome flips twice in two consecutive rounds, the pair (in time) of syndromes is ignored since it most likely indicates a measurement error.
	\item If a pair (in space) of neighboring syndromes flips in the same round, a correction on the data qubit between the pair is applied.
	\item If a syndrome flips in round $r-1$ and its neighboring syndrome flips in round $r$, a correction on the data qubit between the pair (in time) is applied. 
	\item If a syndrome flips only once and in a round other than the last, a correction is applied to a data qubit on the boundary such that the data qubit is not between two stabilizers that did not indicate a syndrome.
	\item If a single syndrome flips only once and in the last round, the information is kept until the next round of error correction. No correction based on this syndrome is applied. In this case the location of the error, if any, is inconclusive without another round of syndrome measurements.
\end{enumerate}
%Rule 1 accounts for measurement errors and reduces the likelihood of mistakingly adding a single data-qubit error as in the earlier example.

%Since fewer errors are more likely, 
We decode by checking the above rules in order and determining the set of data-qubit error locations.
We then switch the order of rules 2 and 3 and determine another set of possible error locations. 
We correct based on the set with fewer error locations, since fewer errors are more likely.
Here we assume that $r=3$.

These rules are equivalent to the minimum weight perfect matching algorithm applied to only neighboring-syndrome pairs, with uniform weight for the same distance.  
Since our surface codes are small, performance of the code does not improve when decoding considers more distant pairs.

We encode these rules into a lookup table.
The lookup table maps the syndrome volume of measurement flips to a set of probable errors on the data qubits. 
The table requires constant time and $2n$ space, where $n$ is the number of data qubits.

\subsection{Improved Stabilizer Measurement Circuits}

\begin{figure}[tb!] 
	\includegraphics[width=0.4\textwidth]{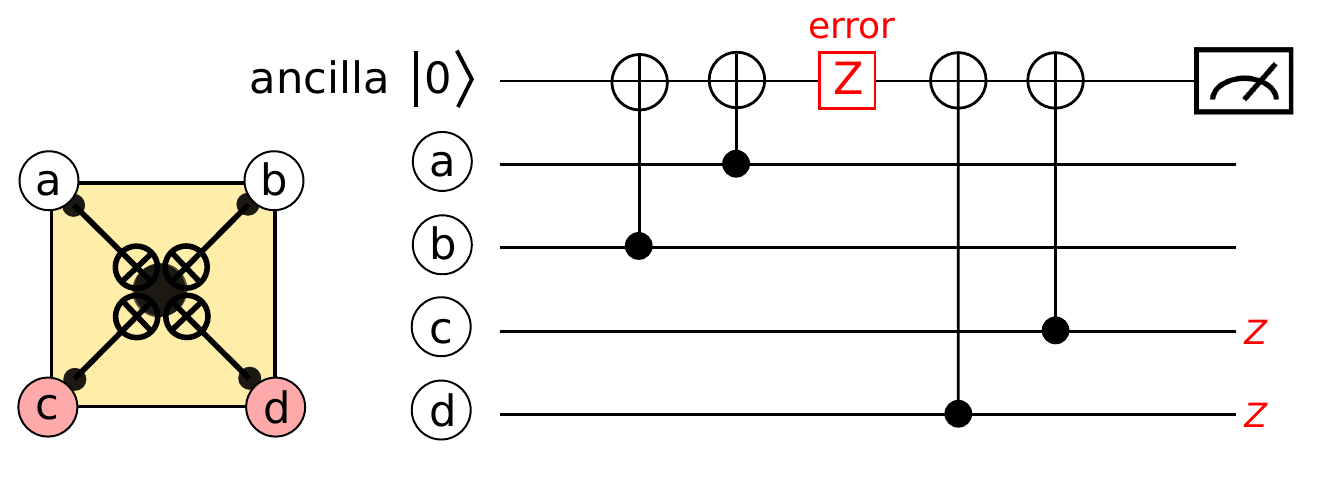} \\ 
	(a)\\
	\includegraphics[width=0.4\textwidth]{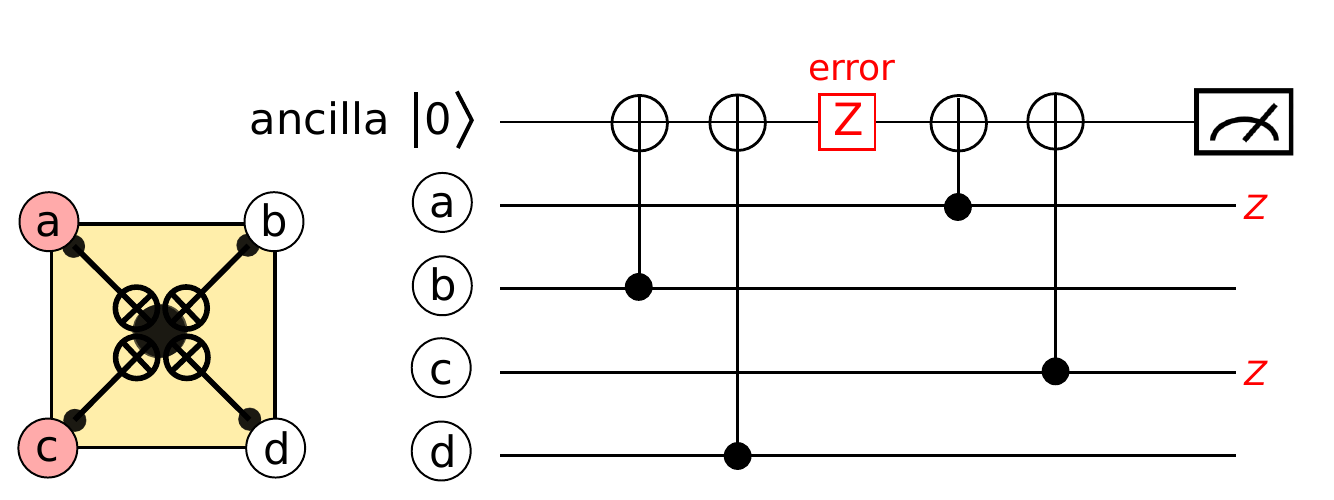} \\
	(b)
	\caption{(a) Stabilizer measurement circuit showing the propagation of a $Z$ error to two $Z$ errors. (b) Reordered stabilizer measurement circuit.}\label{fig_neworder}
\end{figure}

In our simulations of Surface-13 and 17 under noise, we find that using the same CNOT ordering for both $X$- and $Z$-type stabilizer measurements could result in a single error on a syndrome qubit, leading to a logical $X$ or $Z$ error (details on noise are given in Sec.~\ref{sec:noise}).
Figure \ref{fig_neworder}(a) shows an example. 
A $Z$ error on a $Z$-stabilizer syndrome qubit after the first two CNOT gates propagates onto two horizontally aligned data qubits. 
Since our surface codes require only three data qubits to complete a logical error chain, the next round of syndrome measurements will incorrectly diagnose a $Z$ error on the third qubit, leading to a logical $Z$ error chain.

To prevent the creation of a logical error, we propose to measure $X$- and $Z$-type stabilizers in different orders. 
The sequence for $X$ stabilizers is the same as in Figure \ref{fig_oldorder}.
We modify the order of CNOTs in $Z$ stabilizers as: top right ($b$); bottom right ($d$); top left ($a$); bottom left ($c$) (Figure \ref{fig_neworder}(b)). 
This order maintains the alignment of qubits $a$ and $c$ such that they are perpendicular to the direction of the corresponding logical chain.
It also preserves the commutation relations as well as the circuit depth and size. 
Fig.~\ref{fig_neworder} shows an example where two $Z$ errors map to a single $Z$ error with the new order, versus a logical error with the old order. 
%It results in the two $Z$ errors mapping back to a single $Z$ error as opposed to a logical error. 
We use this new order for all simulations in this paper.

%% -------------------------------------------------------------------------------------------------

\section{Noise models}\label{sec:noise}

%% -------------------------------------------------------------------------------------------------

In this section, we present the noise models considered in our surface code simulations.
We review two noise models that can be simulated efficiently on a classical computer (depolarizing and Pauli-twirl approximation) and one noise model that requires exponential memory to simulate (amplitude and phase damping).% and magnetic field noise).

\subsection{Symmetric and Asymmetric Depolarizing Channels}\label{seq_depo}

The depolarizing channel ($D$) is a standard quantum noise model in which a qubit becomes depolarized with a given probability $p$. 
This channel transforms a density matrix of a single qubit as 
\begin{equation}\label{eq:ADC}
\rho \rightarrow \epsilon_{D}(\rho) = p_I \rho + p_X X \rho X + p_Y Y \rho Y + p_Z Z \rho Z,
\end{equation}
where $p_I = (1-p_X-p_Y-p_Z)$. 
In this model, a qubit suffers from discrete Pauli bit-flip ($X$), phase-flip ($Z$), or bit-and-phase flip ($Y$) errors with probabilities $p_X$, $p_Z$, and $p_Y$, respectively. 
When $p_X = p_Y = p_Z$, this channel is called a \textit{symmetric} depolarizing channel.
When the probabilities are independent, the model is called an \textit{asymmetric} depolarizing channel.

%% -------------------------------------------------------------------------------------------------

\subsection{Amplitude and Phase Damping Channel}\label{seq_realdamping}

The amplitude damping channel ($AD$) characterizes the behavior of energy dissipation of the quantum system, including spontaneous emission of a photon from a qubit. 
This channel transforms the density matrix of a single qubit as
\begin{equation}
\rho \rightarrow \epsilon_{AD}(\rho) = E^{AD}_1 \rho E^{AD\dagger}_1  +  E^{AD}_2 \rho E^{AD\dagger}_2,
\end{equation}
where
\begin{equation}
E^{AD}_1 =\begin{pmatrix} 1 & 0 \\ 0 & \sqrt{1-p_{AD}} \end{pmatrix},\, E^{AD}_2 =\begin{pmatrix} 0 & \sqrt{p_{AD}} \\ 0 & 0 \end{pmatrix},
\end{equation}
and $p_{AD}$ is the probability of a qubit emitting a single photon.

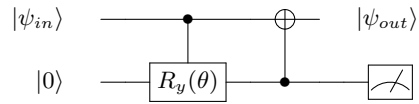
\begin{figure}[tb!]
%\scalebox{0.8}{
\[
\Qcircuit @C=1em @R=1em @!R {
  \lstick{\ket{\psi_{in}}} &  & \qw & \ctrl{1} 			& \qw & \targ 	& \qw   & \rstick{\ket{\psi_{out}}}  & \\
  \lstick{\ket{0}} 	    &   & \qw & \gate{R_y (\theta)}    	& \qw & \ctrl{-1}& \qw  & \qw				& \meter
}
\]
%}
\caption{Circuit representation of amplitude damping~\cite{Nielsen_book}.}
\label{fig_ad}
\end{figure}

Figure~\ref{fig_ad} expresses amplitude damping of a single qubit in the form of a quantum circuit where an ancilla qubit is used to represent the environment and $\sin^2 (\theta/2) = p_{AD}$~\cite{Nielsen_book}.
The input is an arbitrary single-qubit state $\ket{\psi_{in}} = a \ket{0} + b \ket{1}$ and the output state is given by
\[ \ket{\psi_{out}} = \left\{ 
   \begin{array}{l l}
     N a \ket{0} +  N b \sin(\theta/2)\ket{1} & \quad \text{if measure 0}\\
     \ket{0} & \quad \text{if measure 1,}
   \end{array} \right.\]
where $N$ is a normalization constant. The probabilities of measuring 0 and 1 are $1-b^2 p_{AD}$ and $b^2 p_{AD}$, respectively.

During simulation, we do not use an extra ancilla as shown in the circuit in Figure \ref{fig_ad}. Instead, we calculate the probability of measuring $0$ and $1$ given input state $\ket{\psi_{in}}$, and simulate the measurement outcome with a random number. When the simulated measurement is $0$, we apply the rotation $R_y(\theta)$ on $\ket{\psi_{in}}$. When it is $1$, we apply damping and the state becomes $\ket{0}$.

%KMS: ADD DETAIL ON HOW THIS IS IMPLEMENTED IN LIQUID WITHOUT USING AN EXTRA QUBIT. YT: added

The phase damping channel ($PD$) is described similarly as 
\begin{eqnarray}
\rho \rightarrow \epsilon_{PD}(\rho) = E^{PD}_1 \rho E^{PD\dagger}_1  +  E^{PD}_2 \rho E^{PD\dagger}_2,
\end{eqnarray}\label{eq_pd}
where
\begin{equation}
 E^{PD}_1 = \begin{pmatrix} 1 & 0 \\ 0 & \sqrt{1-p_{PD}} \end{pmatrix}, \,
 E^{PD}_2 = \begin{pmatrix} 0 & 0 \\ 0 & \sqrt{p_{PD}} \end{pmatrix}.
\end{equation}

Phase damping noise, also called pure dephasing, is equivalent to the phase-flip channel. 
By unitary freedom of operator-sum representation, we can derive a new set of operation elements to express the channel in terms of the probability of a phase-flip ($Z$) error,
\begin{equation}
 E'^{PD}_1 = \sqrt{1-p_Z} \begin{pmatrix} 1 & 0 \\ 0 & 1 \end{pmatrix}, \, E'^{PD}_2 = \sqrt{p_Z}\begin{pmatrix} 1 & 0 \\ 0 & -1 \end{pmatrix},
\end{equation}
where $p_Z = \frac{1-\sqrt{1-p_{PD}}}{2}$. 

We assume that amplitude and phase damping ($APD$) are the main sources of decoherence. 
Using these two channels together, decoherence on a single qubit transforms the density matrix as
\begin{equation}\label{dephasing}
\rho \rightarrow \epsilon_{APD}(\rho) = \begin{pmatrix} 1-\rho_{11} e^{-t/T_1} & \rho_{01} e^{-t/T_2} \\ \rho^{*}_{01} e^{-t/T_2} & \rho_{11} e^{-t/T_1}\end{pmatrix},
\end{equation}
where $t$ is the execution time of the gate including identity, $T_1$ and $T_2$ are the single-qubit relaxation and dephasing times, respectively, and $e^{-t/T1} = 1-p_{AD}$ and $ e^{-t/T2} = \sqrt{(1-p_{AD})(1-p_{PD})}$~\cite{Nielsen_book}.
%
%Table \ref{damping_blochs} shows the Bloch sphere representation of amplitude damping channel. 
%
%\begin{table}[h!] \centering
%\begin{tabular}{c c c} 
% \includegraphics[width=0.3\textwidth]{blochspheres/plain.pdf} 
%& \includegraphics[width=0.3\textwidth]{blochspheres/ampdamp.pdf}
% & \includegraphics[width=0.3\textwidth]{blochspheres/ampdamp_side.pdf} \\
%plain Bloch sphere & amplitude damping & amplitude damping (side view)  \\
%\end{tabular}\caption{Amplitude damping channel ($p_{AD} = 0.6$) represented using Bloch sphere.}\label{damping_blochs}
%\end{table}

%% -------------------------------------------------------------------------------------------------

\subsection{Approximate Amplitude and Phase Damping Channel}

Using a technique called Pauli twirling ($PT$) \cite{Silva2008}, a Pauli channel $\epsilon_T$ can be used to approximate the decoherence channel given in Eq~\ref{dephasing} \cite{Ghosh2012, Sarvepalli2009}, where 
\begin{eqnarray}
\epsilon_{PT} (\rho) &=& \frac{1}{4} \sum_{A \in {1-X,Y,Z}} A^\dagger \epsilon(A \rho A^\dagger) A.
%&=& \frac{1}{4} \left ( X\epsilon (X \rho X ) X + Y \epsilon (Y \rho Y) Y + Z \epsilon (Z \rho Z) Z  \right ).
\end{eqnarray}
Twirling results in removal of the off-diagonal terms and in turn allows expression of the channel as an asymmetric depolarizing noise channel (given in Eq~\ref{eq:ADC}) with the
%\begin{equation}
%\rho \rightarrow \epsilon_{ADC}(\rho) = p_I \rho + p_X X \rho X + p_Y Y \rho Y + p_Z Z \rho Z,
%\end{equation} 
probabilities given by
\begin{eqnarray}\label{oneqdephasing}
p_X = p_Y = \frac{1-e^{-t/T_1}}{4},   \label{eq_probx}\\
p_Z =  \frac{1-e^{-t/T_2}}{2}-  \frac{1-e^{-t/T_1}}{4},\label{eq_probz}
\end{eqnarray}
where the probabilities of failure are expressed in terms of the execution time $t$ of a gate, the qubit relaxation time $T_1$,  and the qubit dephasing time $T_2$~\cite{Ghosh2012}. 

Assuming errors are independent, the probabilities of two-qubit errors, for example when a CNOT gate fails, are approximated as in \cite{Ghosh2012} as
\begin{eqnarray}\label{twoqdephasing}
p_{I(X or Y)} &=& p_{(X or Y)I} = p_X(1 - p_X - p_Y - p_Z), \nonumber \\
p_{(X or Y)(X or Y)} &=& p_{X} p_X ,\nonumber  \\
p_{Z(X or Y)} &=& p_{(X or Y)Z} = p_X p_Z, \nonumber \\
p_{IZ} &=& p_{ZI} = p_Z (1 - p_X - p_Y - p_Z), \nonumber \\
p_{ZZ} &=& p_Z p_Z.\nonumber 
\end{eqnarray}

\section{Experimental Setup}\label{sec_steps}

We use the \Liquid software architecture \cite{Liquid_Documentation} to perform simulations of the surface code under noise.
\Liquid (Language-integrated Quantum Operations) contains an embedded, domain-specific language for programming quantum circuits as well as two circuit simulation environments.  The first environment allows efficient simulation of Clifford circuits, based on the Gottesman-Knill theorem, and is called Stabilizer simulation \cite{Gottesman1998,Aaronson2004}.
The second environment, called Universal simulation, allows full simulation of arbitrary quantum circuits.

While some of our noise models allow Stabilizer simulation, we have chosen for consistency to perform all simulations within the Universal simulation environment.
\Liquid allows universal simulation of a number of qubits that is limited by the main memory of the machine.
We ran simulations on a large HPC cluster containing several hundred nodes with 32GB of RAM each, allowing simulation of up to roughly 30 qubits on each node.
Our simulations required thousands of hours of compute time.

\subsection{Monte-Carlo Simulation}

%circuits in order to determine the threshold.
%The fault-tolerance threshold is a widely used performance measure of quantum error-correcting codes.
%The threshold value is the error rate below which the circuit operations must operate in order for the encoded circuit to be more protected than the unencoded circuit. KMS:REWORD.
%It guarantees that any quantum circuit encoded with the code can achieve arbitrary accuracy if every quantum gate can be implemented with error rate below the threshold. 

\begin{figure}[tb] 
	\includegraphics[width=0.4\textwidth]{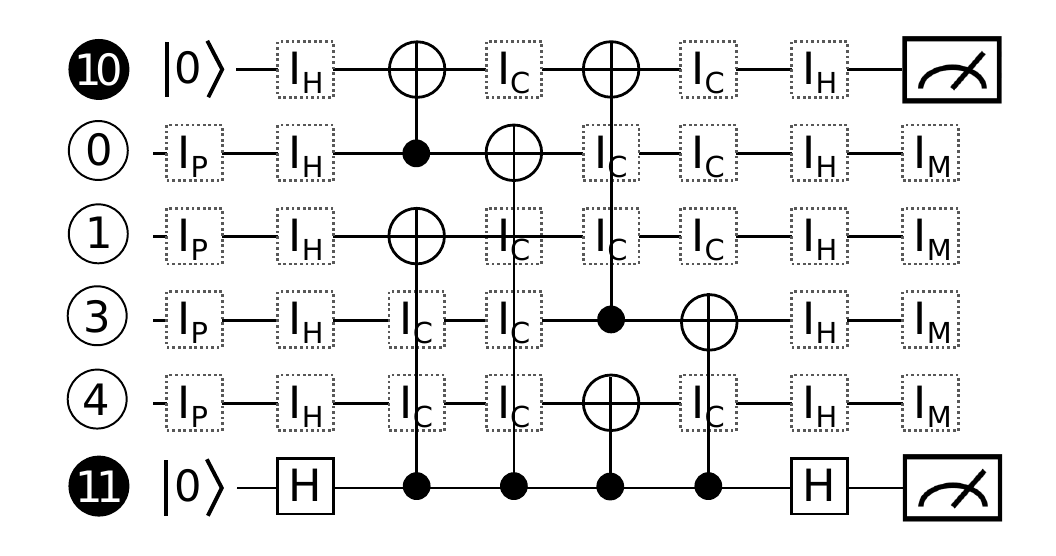}
	\caption{Insertion of identity gates during simulation of the circuit of Fig.~\ref{fig_17_13_circ}(a). Four different identity gates are used based on the other location type in the given timestep: prepare (P), single-qubit gate (H), two-qubit gate (C), and measurement (M). }\label{fig_circ_identity}
\end{figure}

%KMS: SHOW FIG 2 WITH IDENTITIES AND TIMESTEP LINES AS EXAMPLE
We restrict the operations in our circuits to the five types given in Table \ref{tab_counts}, which we refer to as \textit{location types}: $I$, $H$, CNOT, Prepare a $\ket{0}$ state, and Measure in the $Z$ basis. 
When no location type is specified on a qubit, the identity gate $I$ is applied to that qubit, where the duration of the identity is set by the location type occurring on other qubits in the time step.
When a qubit is idle for a duration of $t$ time steps (while gates are being applied on other qubits), we apply $t$ identity gates to it to simplify the simulations. 
Figure \ref{fig_circ_identity} shows the circuit of Figure \ref{fig_17_13_circ}(a) with identity gates inserted. 
%KMS: MORE DETAIL OR CIRCUIT DIAGRAM EXAMPLE? GIVE AS PART OF ABOVE FIG? YT: figure added. 
Further circuit optimization can be performed, for example by delaying qubit preparation and measuring a qubit as soon as gate operations complete.
Such optimization will result in improved thresholds.  For simplicity, we choose to maintain gate alignment between stabilizers.

We perform Monte-Carlo simulation of the surface code layouts to compute the logical error rates.
At each time step of the circuit, each qubit undergoes a location type followed by the given noise model.
For depolarizing noise and approximate damping noise, we replace each location type except measurement by the location type followed by an $X$, $Y$, or $Z$ gate (``error") with probability $p_X$, $p_Y$, and $p_Z$, respectively. 
In the case of measurements,  $X$, $Y$, or $Z$ errors are placed before the measurement location.
%KMS: HOW ARE THE PROBABILITIES SET?  ALL EQUAL?  WHAT ABOUT TWO QUBIT?  DEPO ASSUMED AS EQUAL FOR EACH?

For amplitude and phase damping and the Pauli-twirl approximation,
% and uniform magnetic field noise
we apply the noise model after every location given the duration of the current time step $t$. 
%Approxima and real decoherence are applied after each location with probability dependent on the location duration.
The duration values we consider are given in Table \ref{params} of Section \ref{sec_architecture}.

%of the three distance-3 surface codes which could serve as the expected values of the experiments following the steps described in Section \ref{sec_steps}. 
%% -------------------------------------------------------------------------------------------------

\subsection{Logical Error Rate Calculation}

\begin{figure}[tb]
	\includegraphics[width=0.4\textwidth]{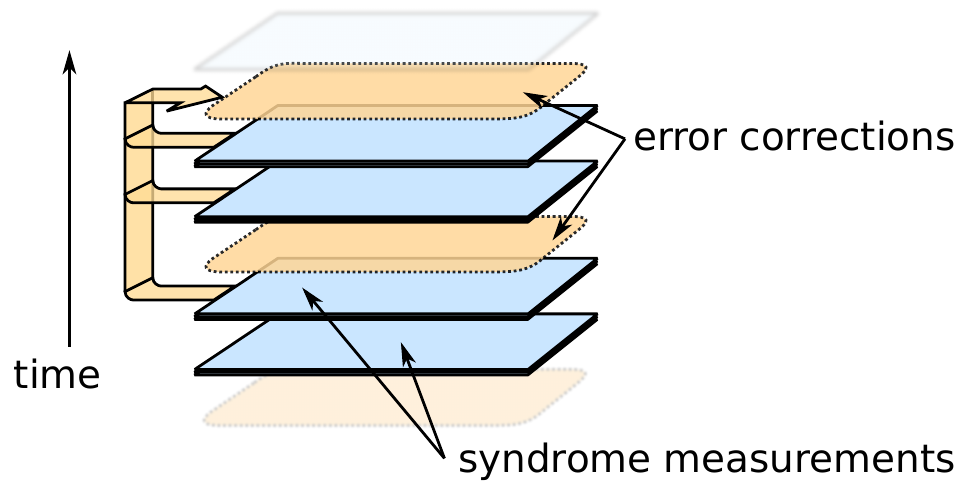} \\ 
	\caption{Illustration of the syndrome volume.  Each window consists of three rounds of the surface code. Corrections are applied to the state after the final round in the window.  The state is then checked for a logical error.  An example window consists of the top three blue layers, where one layer overlaps with the previous window.}\label{fig_layers}
\end{figure}

We calculate the logical error rate of a given layout by simulating it under the various noise models.
%Combining the encoding and the decoding scheme described above, we run full simulations of the three distance-3 surface codes to obtain logical error rates of these codes under various noise models. 
At the start of each simulation, we initialize all data qubits to $\ket{0}$ (if preparing $\ket{0_L}$) or $\ket{+}$ (if preparing $\ket{1_L}$) and run a noise-free cycle of syndrome measurements to project into an initial stabilizer state of the code. 
We refer to this state as the \textit{quiescent} state \cite{Fowler092012}.
Note that for a code with $s$ stabilizers, there are $2^s$ possible quiescent states, since each stabilizer measurement can randomly project to either a $\pm 1$ eigenstate.
In the absence of noise, the quiescent state will be maintained during subsequent rounds of the surface code.

After initialization of the quiescent state, the simulation proceeds as follows:
\begin{enumerate}
\itemsep 1pt
\parskip 0pt
\item Execute two rounds of the surface code with noise (execute three if this is the first execution of the loop). Record the list of syndrome flips between contiguous rounds in the syndrome volume. For the first round, compare to the quiescent state.
\item Apply the decoder (Section \ref{sec_decoding}) to the three-layer syndrome volume to determine the most probable set of error locations.  
\item Apply noise-free corrections to the state. In practice, corrections can be tracked directly in software.
\item Check for a logical error by calculating the distance of the state to the possible logical states.  If the closest logical state is incorrect, count a logical error.
\item Repeat from Step 1 until $m$ logical errors are detected.
\end{enumerate}

After each logical error check (Step 4) the syndrome volume contains a list of unpaired syndrome flips due to the last two rules of our decoder. 
Each syndrome volume, as shown in Fig.~\ref{fig_layers}, thus contains three layers: the final layer from the previous volume and two layers from two additional rounds of the surface code.
We refer to the number of rounds in the volume as the \textit{window size}.  
We experimented with various window sizes and found three was optimal for distance-three layouts. 
%Each syndrome volume always contains three rounds of the surface code.
In our simulations, $m$ varies between 10 and 200 depending on the size of the physical error rates.

%Figure \ref{fig_layers} illustrates rounds of the surface code interspersed with corrections and logical error checks.
%We then repeat the whole cycle by applying the syndrome measurements twice, determine error locations, correct, and check for a logical error, until we observe $n$ logical errors. 
%In our simulation, we correct errors based on the previous three rounds of syndrome measurements.

We calculate the logical error rate per window since in an experiment, the logical qubit will be measured after completion of a window to ensure optimal decoding and correction.
For a window containing $r$ rounds, the logical error rate $P_r$ is given by 
\begin{equation}
\label{eq_fail}
P_r = m / R,
\end{equation}
where $R$ represents the number of windows executed to observe $m$ logical errors. 
%Logical errors are detected at the completion of each window.
%KMS: HOW MANY RUNS? added two paragraphs above (10-200)
When $r=1$, Eq \ref{eq_fail} represents the logical error rate per round of the surface code.

Since we only calculate $P_r$ for distance $d=3$, we estimate the pseudothreshold \cite{Svore05,Svore05b}, denoted as $P_r^{th}$ as opposed to the asymptotic threshold as $d \rightarrow \infty$.
The pseudothreshold can be defined by the crossing point between the line $x=y$ and the plot $p$ vs.~$P_r$.
If the error rate $p$ of each physical location type falls below the pseudothreshold $P_r^{th}$, then the code is guaranteed to lower the logical error rate below $p$.
%For the given layout, this pseudothreshold represents the error rate below which the location types must operate in order to decrease the logical error rate.
%KMS: REWORD THIS SENTENCE 
%KMS: IS THIS THE CORRECT FORMULA WE ARE USING? YT: Removed r from the equation.

%Since we decode over a three-round window, we also calculate the logical error rate per window, which we express as the logical error rate per $r=3$ rounds using the formula above.
%\begin{equation}
%P_{3} = n/R.
%\end{equation}
%where $W$ represents the number of windows required for $n$ logical errors to appear, where in our case $W=R/3$.
%

The logical error rate per window $P_r$ and the logical error rate per round $P_1$ are related by 
\begin{equation}\label{eqn_pw}
P_r \approx r P_1(P_1)^{r-1}+(r-2)P_1^3 (1-P_1)^{r-3}.
\end{equation}

For depolarizing noise, we calculate $P_1$ (to compare with previous work) and $P_{3}$. 
For amplitude and phase damping and the Pauli-twirl approximation, we calculate $P_{3}$.%, which represents the logical error rate during an actual experiment, based on the decoder of Section \ref{sec_steps}. 
 
%% -------------------------------------------------------------------------------------------------

\subsection{Architectural Settings}\label{sec_architecture}

%% -------------------------------------------------------------------------------------------------
% * assumed the same to be measurement
%
%sc is from the emails of Krysta and Leo DiCarlo
%
%Iontrap T1/T2 ratio from http://arxiv.org/pdf/1303.5770.pdf
%
%MUSIQC time from http://arxiv.org/pdf/1208.0391.pdf
%
%Check iontrap (now) time from Ken's email. Check Moroe (rotations) and Jungsang's (measurements) papers for references.
%% -------------------------------------------------------------------------------------------------
\begin{table*}[tb] 
\caption{Qubit relaxation, dephasing, and gate times assumed for different architectures. DiVincenzo and Helmer parameters are taken from \cite{Ghosh2012}. $SC$ denotes superconductor; $IT$ denotes ion trap architecture.}\label{params}
%\scalebox{0.7}{
\begin{tabular}{| c | c | c | c | c |c|c|c|}
\hline
Parameter 	& Description/Location			&$SC_S$ (Slow)		&  $SC_F$ (Fast) 		& $SC_D$ (DiVincenzo) & $SC_H$ (Helmer) 		 	& $IT_S$ (Slow) 		& $IT_F$ (Fast) \\  \hline
$T_1$		& qubit relaxation time			& $T_1$				& $T_1$	  & $T_1$			&	$T_1$		&  $T_1$			& $T_1$  	\\ \hline
$T_2$		& qubit dephasing time			& $T_1$			& $T_1$			& $2\ T_1$ &   $T_1$					&  $0.1 \ T_1$			& $0.1 \ T_1 $		\\ \hline
$t_{\mbox{prep}}$	& state preparation		& 5 $\mu$s	   	& 1 $\mu$s & 40 ns				& 40 ns	& 100 $\mu$s		& 30 $\mu$s	\\ \hline
$t_{1}$	& single-qubit rotation 			& 100 ns	  	& 10 ns & 5 ns				& 5 ns					& 1    $\mu$s		& 1 $\mu$s 	\\ \hline
$t_{\mbox{meas}}$	& measurement			& 5 $\mu$s	   	& 1 $\mu$s & 35 ns			& 35 ns			   	& 100 $\mu$s		& 30 $\mu$s		\\ \hline
$t_{\mbox{CNOT}}$	& CNOT					& 1 $\mu$s	   	& 100 ns & 80 ns & 20 ns					  	& 100 $\mu$s		& 10 $\mu$s		\\ \hline
$t_{r, 13}$			& one round (S-13)		& 28.2 $\mu$s &  4.82 $\mu$s &  800 ns &  320 ns   &  1202 $\mu$s &  202 $\mu$s\\ \hline
$t_{r, 17 \& 25}$	& one round (S-17, S-25) &  14.2 $\mu$s &  2.42 $\mu$s &  405 ns & 165 ns   &  602 $\mu$s &  102 $\mu$s  \\ \hline
\end{tabular}
%}
\end{table*}

For amplitude and phase damping and the Pauli-twirl approximation, we consider several parameter settings derived from superconductor and ion trap architectures. 
These architectures are well-suited to 2-D, nearest-neighbor operations required for the surface code.
Table \ref{params} lists the different parameter settings considered for each architecture. 
%KMS: TEXTBOOK and HELMER LOOK IDENTICAL. WHAT ARE VALUES USED FOR T1?  YT: The only value different between texbook and helmer are CNOT gates (21ns and 20ns) T1 value is a main variable which we used the same range. In Ghosh's work, they used different pCX value (probability of stochastic error after CX) which was added to the amplitude damping noise. We decided to only see amplitude damping noise and set all other stochastic error rates (measurement and prep errors) zero. Should we use only one of them?
%KMS: TEXTBOOK ARE VERY SHORT COMPARED TO SC NOW. WHY?
The time per round $t_{r,\{13,17,25\}}$ indicates the time required to complete one round of the surface code given the other parameters.  
These six architecture settings represent a range of round times between $165$ ns to $602\times10^3$ ns for Surface-17 and Surface-25.  
Note that the Surface-13 layout requires roughly twice the amount of time of Surface-17.

2D superconducting architectures have demonstrated fast single- and two-qubit gate execution times in recent years \cite{DiCarlo10,Chow10,Chow13,Martinis14}.
Current gate times are in the range of $10$--$20$ ns and $30$--$80$ ns for single-qubit and two-qubit gates, respectively, with experimental $T_1$ times as long as $20$--$40$ $\mu$s \cite{Chow13,Martinis14}.
The  DiVincenzo ($SC_D$) \cite{Divincenzo2009} and Helmer ($SC_H$) \cite{Helmer2009} superconductor parameters are derived from \cite{Ghosh2012}.
$SC_D$ requires longer CNOT gate times than $SC_H$.
$SC_S$ and $SC_F$ represent parameters for slow and fast gate times, respectively, based on recent experiments \cite{DiCarlo10,Chow10}.
In particular, they account for $\mu$s preparation and measurement times, while $SC_D$ and $SC_H$ assume ns times.
%With these four architectural settings, we can determine 
%KMS: ADD HERE BASED ON EXPTS.  
%We focus on the influence of amplitude and phase damping noise and assume that there is no other types of noise including random measurement noise.

Ion traps are another promising architecture with demonstrable quantum gates \cite{KMW,monroe2014,TrueNJP2011,wright2012,Harty14}. 
While trapped ion devices tend to have longer gate execution times than superconductor devices, they have been shown to have much longer relaxation and dephasing times in the range of 780--1800 ms \cite{Haffner2008,Noek2013}.
Recently, a $T_2^*$ time of 50 s has been reported \cite{Harty14}. 
$IT_S$ accounts for gate times observed in current experiments and longer preparation and measurement times \cite{Noek2013,Haffner2008,Ken_email}. 
$IT_F$ accounts for gate, preparation, and measurement times of a proposed scalable ion trap quantum computer model \cite{monroe2014}. 
It assumes that all gate operations are within one Elementary Logic Unit (ELU) with 10--100 qubits arranged linearly. 
ELUs are connected to each other using photonic quantum channels to achieve modular scalability.
%It assumes that all gate operations are within one Elementary Logic Unit of the model.
%Ion trap parameters are collected from [http://arxiv.org/pdf/1303.5770.pdf, http://arxiv.org/pdf/1208.0391.pdf]
%KMS: WHAT DOES THAT MEAN? YT: added description

%% -------------------------------------------------------------------------------------------------
\section{Experimental Results}\label{sec:results}

In this section we analyze numerical Monte-Carlo simulations of the distance-three surface code layouts under the multi-parameter noise models.
We first determine the distance-three layout that admits the highest pseudothreshold under depolarizing noise.
We then study the performance of the preferred layout under several realistic noise models.
In particular, for the six architectural settings we compare the accuracy of the approximate amplitude and phase damping channel, which can be efficiently simulated, to the amplitude and phase damping channel, which requires universal simulation.
In each plot, error bars indicate the upper bound statistical significance using the standard deviation.
%Finally, we evaluate the performance of the Surface-17 layout under magnetic noise for the proposed parameter settings.

\subsection{Depolarizing Noise}

We begin by calculating the symmetric depolarizing noise threshold for each distance-three layout.
In this model, each location fails with probability $p$.
For single-qubit locations, $P_I = 1-p$ and $P_X = P_Y= P_Z= p/3$.
For two-qubit locations, $P_{I,I} = 1-p$ and $P_{\{I,X,Y,Z\},\{I,X,Y,Z\}} = p/15$.
Since the circuits and round times differ, we expect the pseudothreshold to vary for each layout.
\begin{figure}[tb] %\centering
\begin{tabular}{c}
\includegraphics[width=0.4\textwidth]{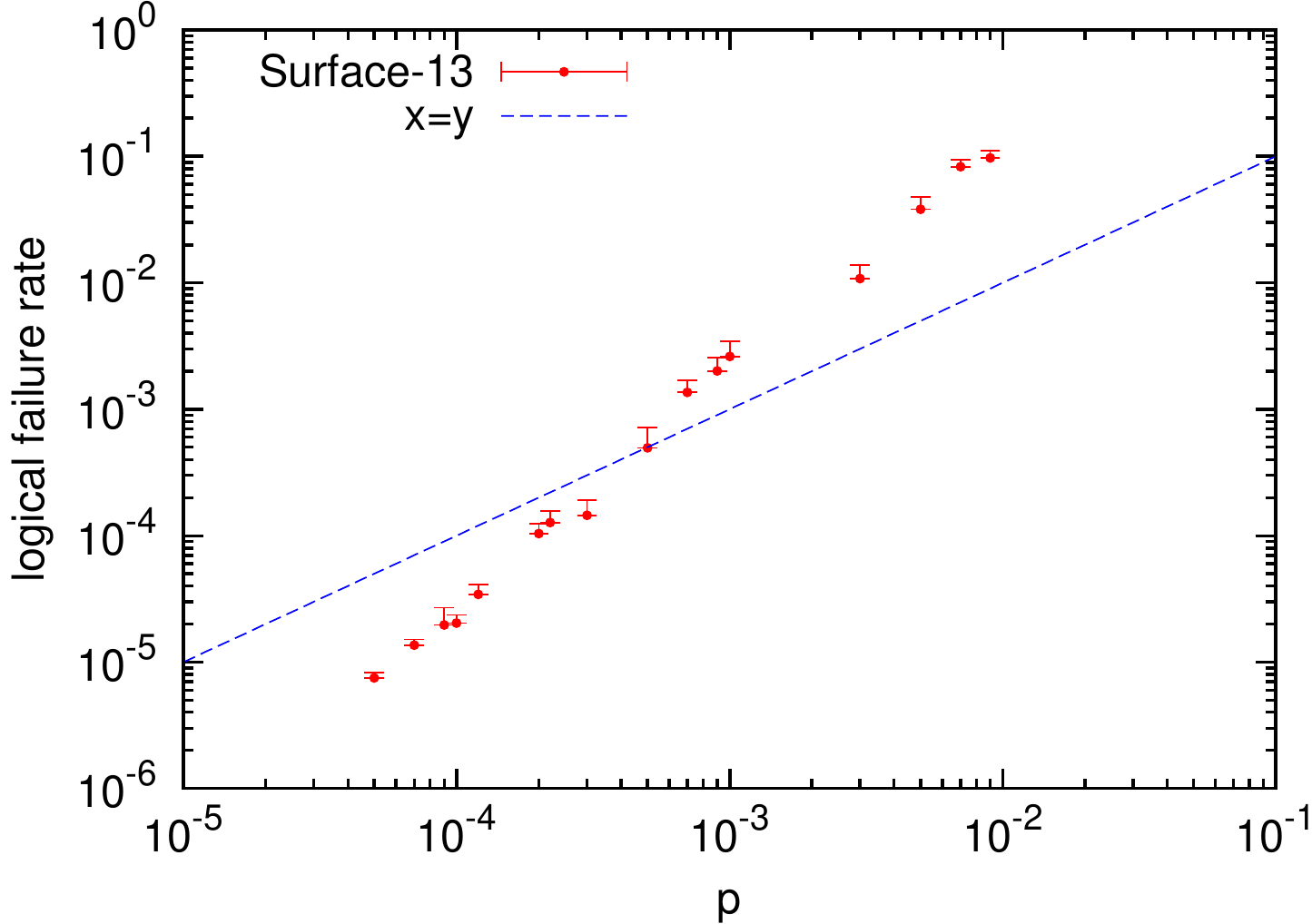} \\
	(a) Surface-13  \\
\includegraphics[width=0.4\textwidth]{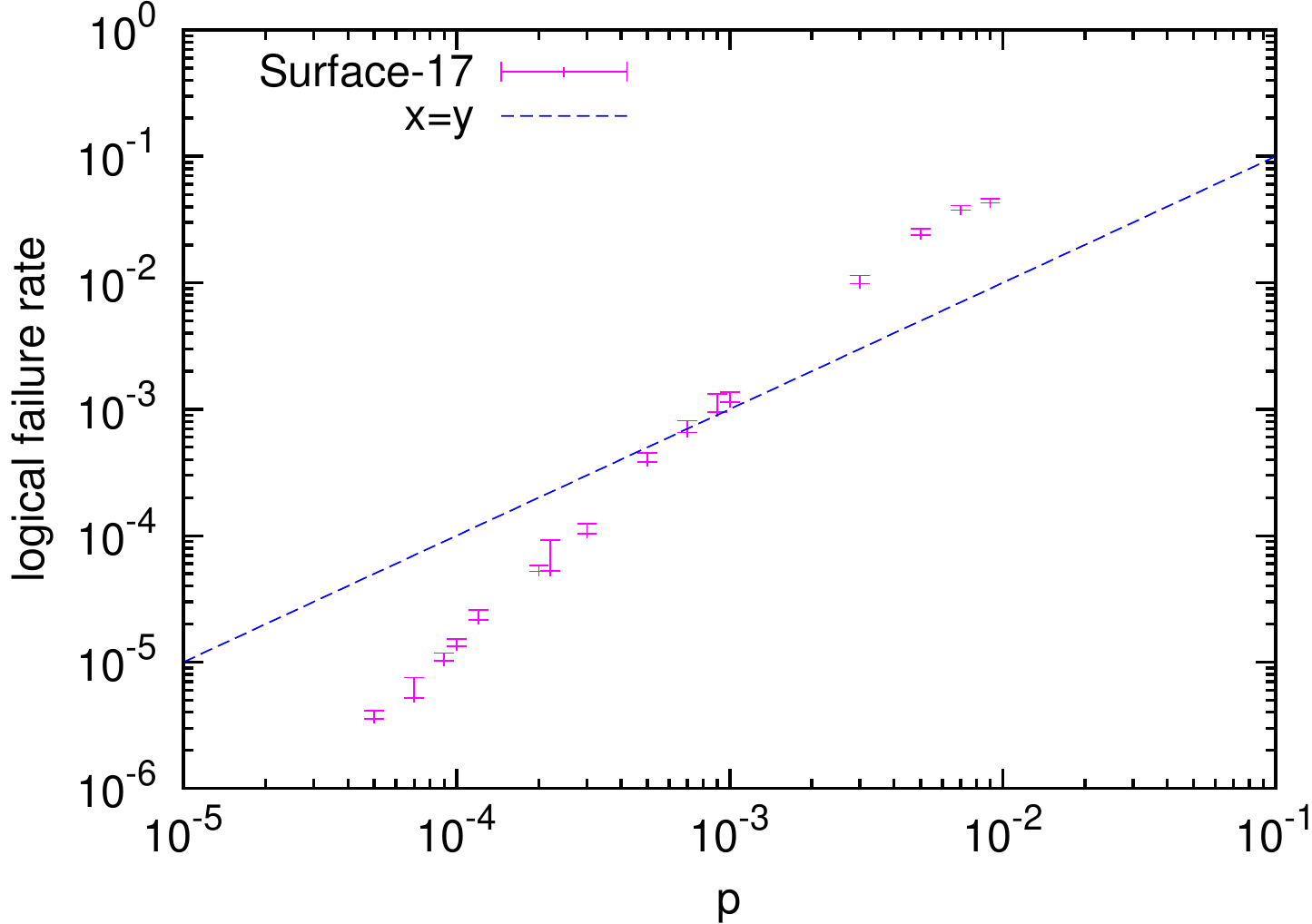}\\
(b) Surface-17	\\
\includegraphics[width=0.4\textwidth]{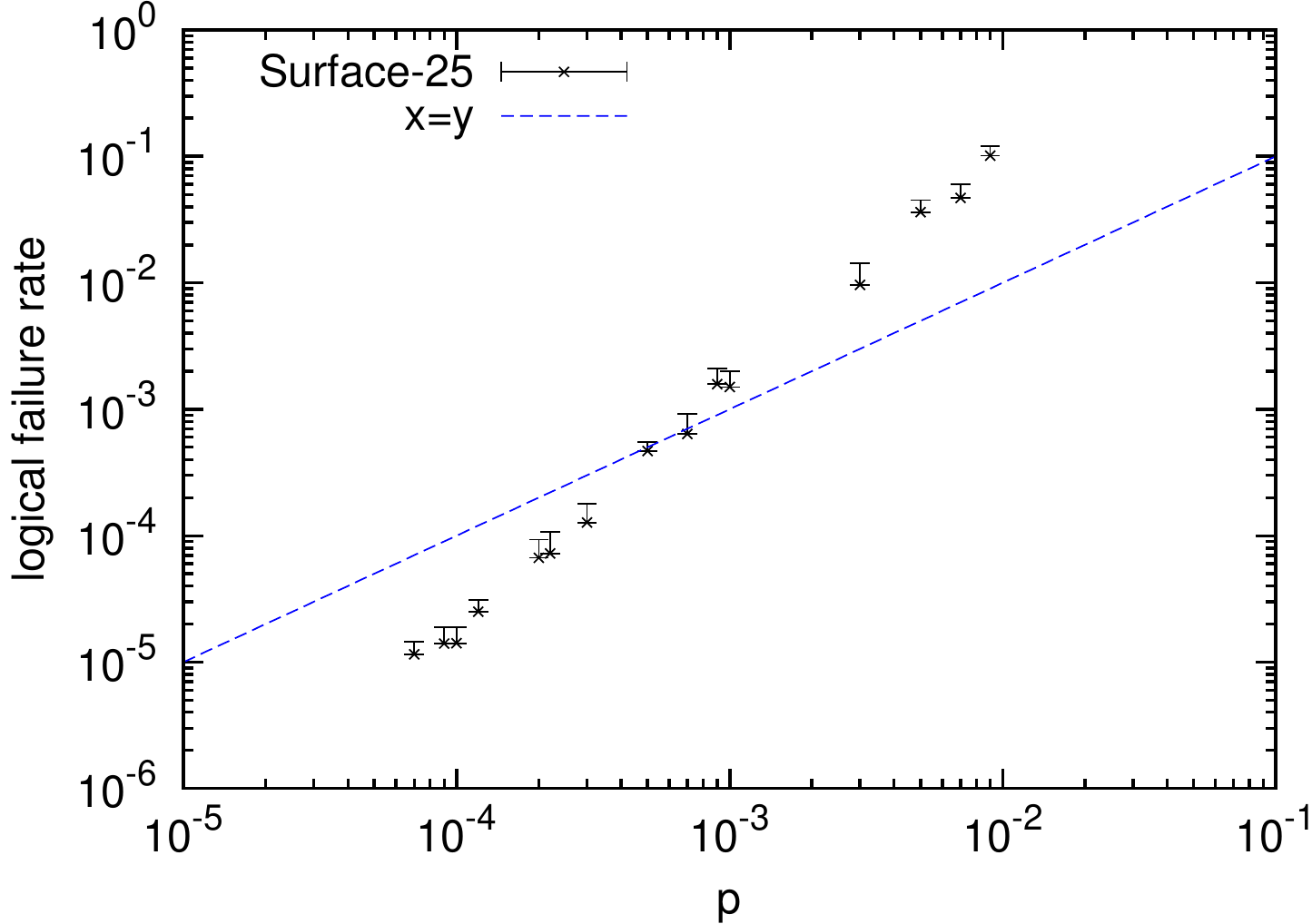} \\
(c) Surface-25
\end{tabular}
\caption{Location error rate $p$ versus logical error rate $P_{1,X}$ for a logical $\ket{1_L}$ state encoded in (a) Surface-13, (b) 17, and (c) 25 under symmetric depolarizing noise. }\label{fig_all_depo}
\end{figure}

Figure \ref{fig_all_depo} plots the location error rate $p$ versus the logical $X$ error rate per round $P_{1,X}$ for Surface-13, 17, and 25, 
where each layout encodes a logical $\ket{1_L}$ state and we check for a logical bit-flip $X_L$.
%KMS: WHY 1 AND NOT 0 STATE? YT: used the same starting state as decoherence
%KMS: MAKE SAME Y AXIS FOR EACH PLOT XXX
Each point represents between $10$ and $200$ independent simulation runs. 

The corresponding pseudothresholds calculated per round ($P^{th}_{1,X}$) and per window ($P^{th}_{3,X}$) are given in Table \ref{tab_thresholds}. 
We find that Surface-13 exhibits slightly lower pseudothresholds due to its higher circuit depth.
Similarly, Surface-25 requires more data qubits and syndrome measurements, thus exhibiting a small decrease in its pseudothreshold as compared to Surface-17.

\begin{table}[tb] 
\caption{Comparison of thresholds and pseudothresholds for the surface code under symmetric depolarizing noise.}
\label{tab_thresholds}
\scalebox{1.0}{
\begin{tabular}{| c | c | c | c |}
\hline
Code	& Threshold & $P^{th}_{1,X}$ & $P^{th}_{3, X}$  \\ \hline
Surface-13	& -	& $3.0 \times 10^{-4}$ & $1.2 \times 10^{-4}$ \\ 
Surface-17  & - & $8.0 \times 10^{-4}$ & $2.0 \times 10^{-4}$ \\  
Surface-25  & - & $5.0 \times 10^{-4}$ & $1.4 \times 10^{-4}$ \\ \hline 
Wang (2011) \cite{Wang2011}  & $1 \times 10^{-2}$ &  -  &  - \\ \hline
Fowler (2012) \cite{Fowler052012,Fowler052012arxiv}  & $9 \times 10^{-3}$ & $\sim 2 \times 10 ^{-3}$ &  - \\ \hline
\end{tabular}
}
\end{table}

Table \ref{tab_thresholds} also contains the pseudothreshold and threshold calculated by Fowler et al.~for Surface-25 \cite{Fowler052012,Fowler052012arxiv}.
Our Surface-25 per-round pseudothreshold is slightly lower.
% than in \cite{Fowler052012}.
Our simulations use a constant window size (Section \ref{sec_steps}) to set the volume history, while Fowler et al.~use a volume including a history of rounds limited only by the data available.
They perform minimum weight matching continuously, round by round, based on a large volume, while we perform correction based on our lookup table and three rounds of history in the volume. 
% the maximum the window size for each value of $p$.
%They sweep over a range of window sizes from $1$ to $25780$ and for each $p$ between $0.05$ and $0.0001$.
We use a static, small window in order to mimic future experimental implementations which are likely to be limited to a small number of rounds and to restricted cold classical processing.
%While it has been shown that a smaller window can lead to some logical error patterns not present when using a larger window \cite{Fowler052012,Fowler052012arxiv},
%We find for window size $3$, our per-round pseudothreshold closely matches that in \cite{Fowler052012,Fowler052012arxiv}.

\begin{figure}[tb] %\centering
	\includegraphics[width=0.4\textwidth]{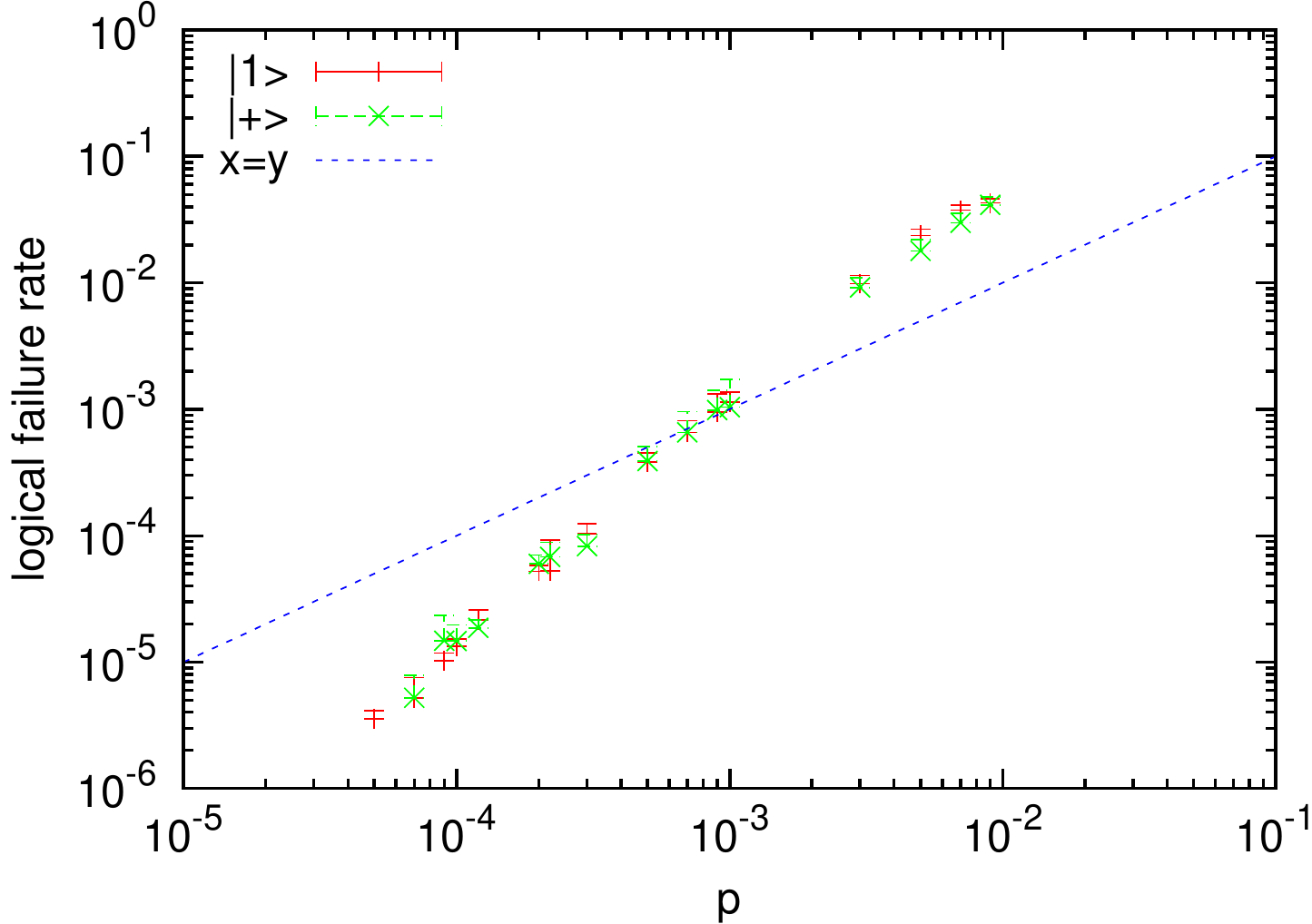}
	\caption{Surface-17 logical $X$ error rate $P_{1,X}$ (red; qubit encoded in $\ket{1_L}$) and logical $Z$ error rate $P_{1,Z}$ (green; qubit encoded $\ket{+_L}$) under depolarizing noise.}
\label{fig_surface17_depo}
\end{figure}

We also calculate the logical $Z$ error rate $P_{L,Z}$ for each layout by encoding a logical $\ket{+_L}$ state and checking for a logical phase flip $Z_L$.
Figure \ref{fig_surface17_depo} plots the location error rate $p$ versus $P_{1,\{X,Z\}}$ for Surface-17.
It is apparent from the plot that the pseudothresholds $P^{th}_{1,X}$ and $P^{th}_{1,Z}$ are comparable.
We find similar results for Surface-13 and Surface-25.

Based on these results, we conclude that Surface-17 is the preferable layout.  
It requires roughly half the depth of Surface-13 and significantly fewer qubits and gates than Surface-25.
In addition, Surface-17 exhibits slightly higher pseudothresholds than the other layouts.
For the remaining experiments, we thus perform all simulations based on the Surface-17 layout.

%% -------------------------------------------------------------------------------------------------

\subsection{Amplitude and Phase Damping vs.~Pauli Twirling}

In this section, we compare the accuracy of the approximate amplitude and phase damping channel using Pauli twirling to the amplitude and phase damping channel.
We first verify that our logical $Z$ and $X$ error rates per round for the Pauli-twirl approximation on Surface-17 align with those reported in \cite{Ghosh2012}.
For $T_1=10$ $\mu$s, we find $P_{Z,1}=4.27\times 10^{-3}$ and $P_{X,1}=4.41\times 10^{-3}$.
These results are very similar to \cite{Ghosh2012}; small differences are expected since Surface-25 is used in \cite{Ghosh2012}.

\begin{figure}[htb] %\centering
	\includegraphics[width=0.4\textwidth]{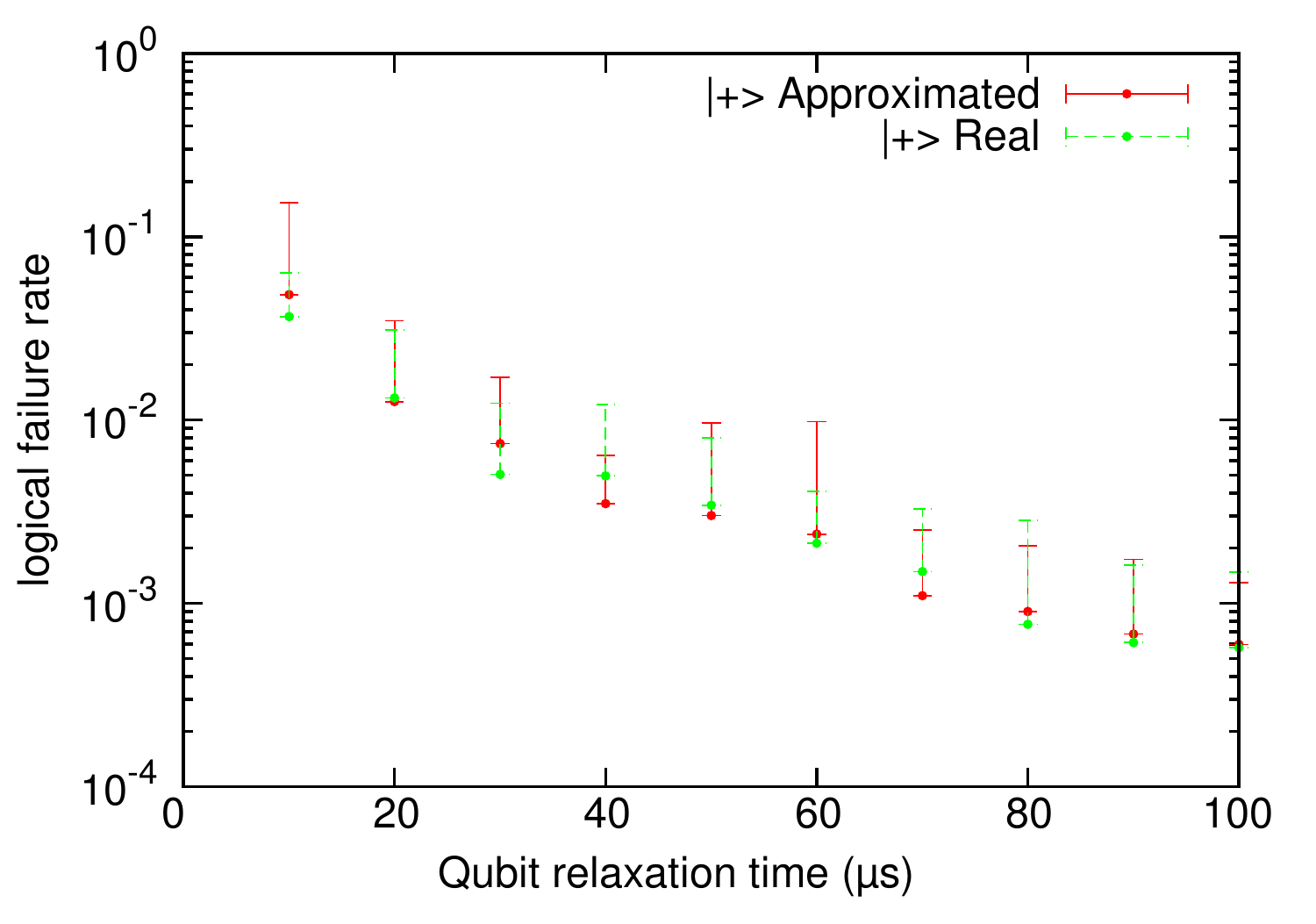}\\
(a) $P_{3,Z}$ \\
	\includegraphics[width=0.4\textwidth]{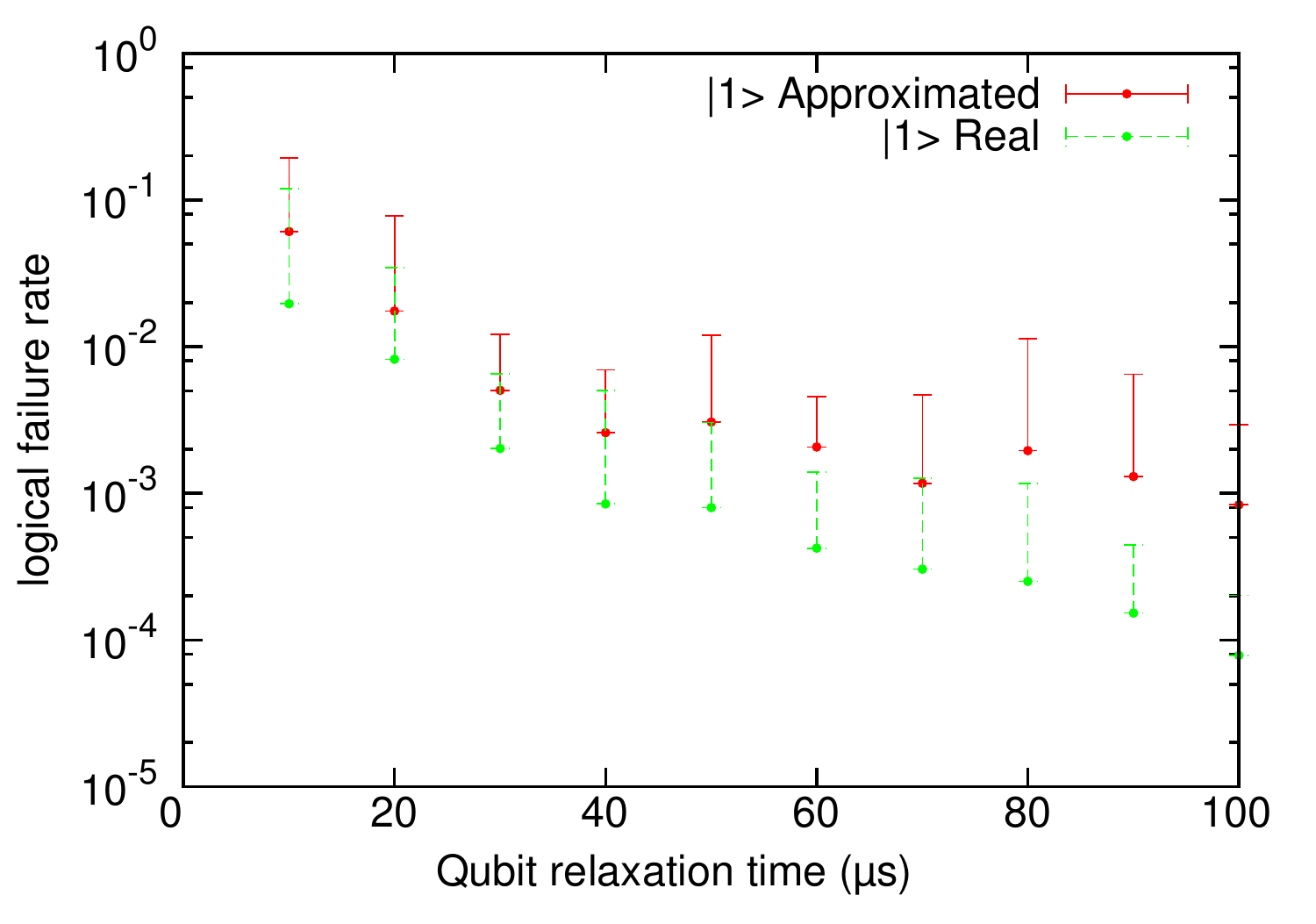}\\
(b) $P_{3,X}$
	\caption{Comparison of the logical error rate of a qubit encoded in Surface-17 under amplitude and phase damping (dashed green) and the Pauli-twirl approximation (solid red) for the $SC_H$ setting. (a) Logical $Z$ error rate $P_{3,Z}$ (on logical $\ket{+_L}$ state); (b) Logical $X$ error rate $P_{3,X}$ (on logical $\ket{1_L}$ state).}
\label{fig_s17_real_vs_pauli}
\end{figure}

We then calculate the logical $Z$ error rate per window, $P_{3,Z}$, for a qubit in the encoded $\ket{+_L}$ state in Surface-17 for both channels for the Helmer setting ($SC_H$).
Figure \ref{fig_s17_real_vs_pauli}(a) plots $T_1$ versus $P_{3,Z}$ for approximate (solid red) and amplitude and phase damping (dashed green). 
We see that the approximate channel using Pauli twirling results in a logical $Z$ error rate that closely matches that of the actual channel.

We also calculate the logical $X$ error rate per window, $P_{3,Z}$, for a qubit in the encoded $\ket{1_L}$ state in Surface-17 for both channels under $SC_H$, plotted in Fig.~\ref{fig_s17_real_vs_pauli}(b).
We find that the approximation channel results in much higher logical $X$ error rates, in particular as the qubit relaxation time $T_1$ increases.
Pauli twirling results in a pessimistic estimate of the error rate, indicating that the threshold under decoherence may be significantly better than previously calculated with this technique.

%% -------------------------------------------------------------------------------------------------

%\subsection{Surface 17 with real decoherence and various architecture settings}

\begin{figure*}[tb!] %\centering

	\includegraphics[width=0.4\textwidth]{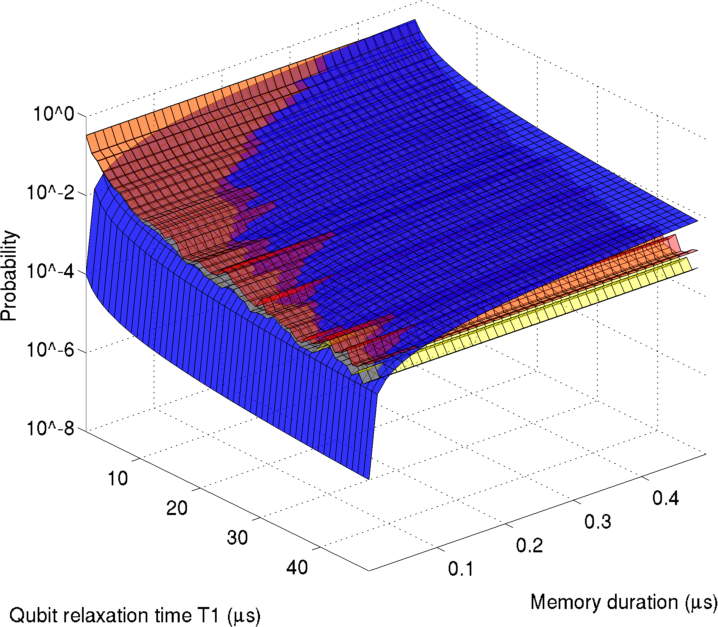}
	\includegraphics[width=0.3\textwidth]{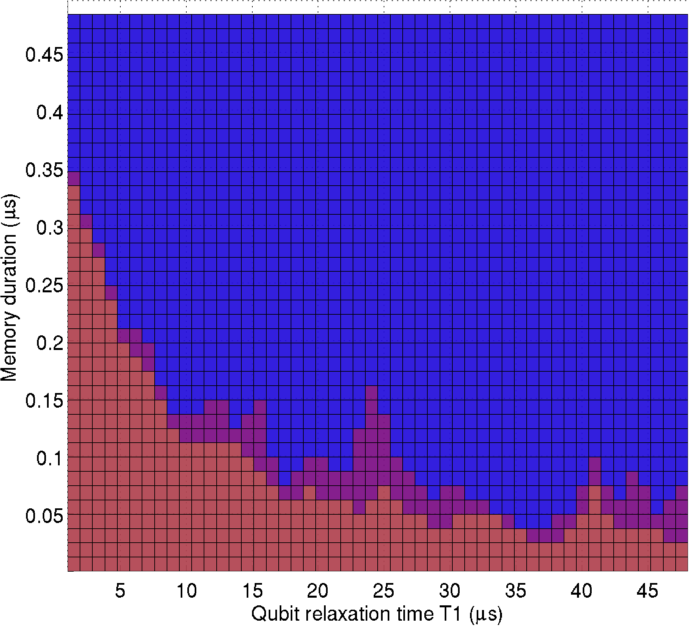}\\	
	(a) Pauli-twirl approximation\\	
	\includegraphics[width=0.4\textwidth]{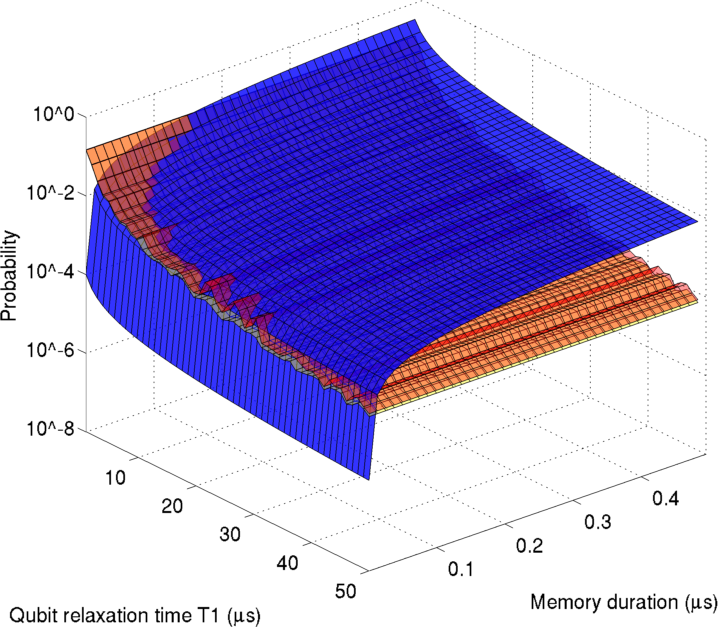}
	\includegraphics[width=0.3\textwidth]{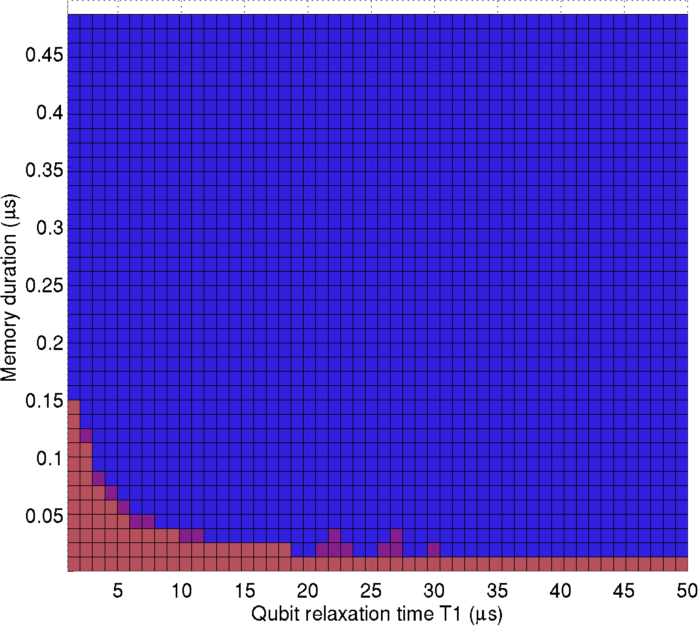}\\
(b) Amplitude and phase damping
	\caption{(Left) 3D plots of $T_1$ time ($\mu$s) vs.~memory duration (ns) vs.~$P_{3,X}$ for the $SC_H$ setting under (a) the Pauli-twirl approximation and (b) amplitude and phase damping. 
The blue surface represents the amplitude damping probability at a given $T_1$ and memory duration (unencoded qubit). 
The yellow surface is the simulated logical error rate given $T_1$ for a qubit encoded in the $\ket{1_L}$ state in Surface-17. 
The orange surface indicates the upper error bar on $P_{3,X}$. 
(Right) 2D plots from the $+z$-axis. 
The blue and red regions indicate a range of $T_1$ times ($x$-axis) for which encoding a qubit in the $\ket{1_L}$ state in Surface-17 reduces or increases, respectively, the logical error rate compared to an unencoded $\ket{1}$ qubit in memory for a range durations ($y$-axis). 
%The red region indicates a range of $T1$ times for which the encoding the qubit will lower the error rate compared to an unencoded qubit (that the qubit is expected to have lower error rate by leaving it idle than when encoded using Surface-17.
}
\label{fig_s17_surfaces}
\end{figure*}

\begin{figure*}[tb!] %\centering
\begin{tabular}{ c c c }

	\includegraphics[width=0.3\textwidth]{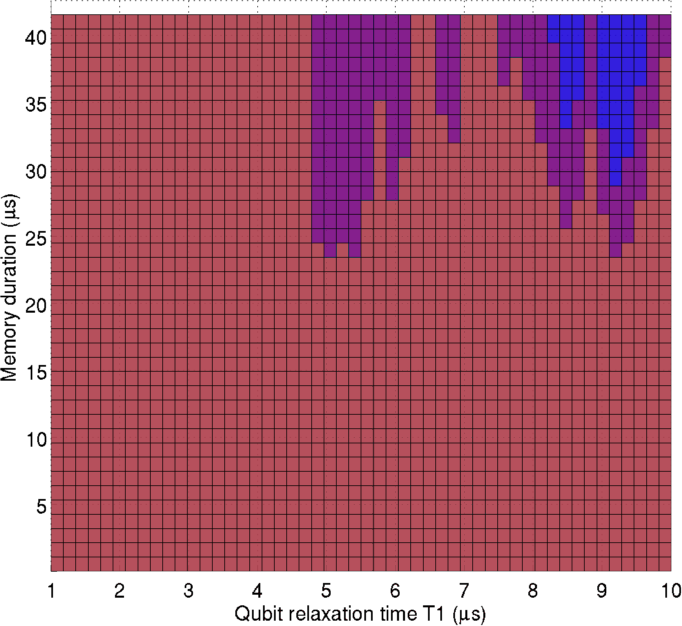}&
	\includegraphics[width=0.3\textwidth]{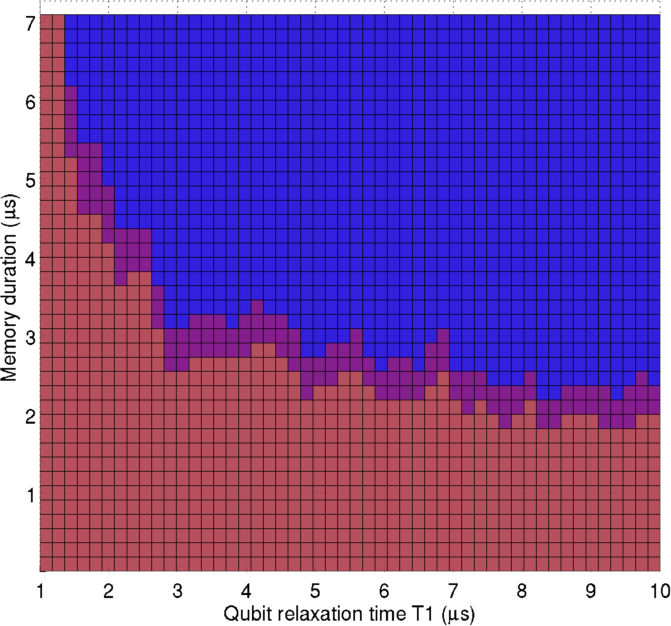} \\	

	(a) $SC_S$ & (b) $SC_F$ \\
	\includegraphics[width=0.3\textwidth]{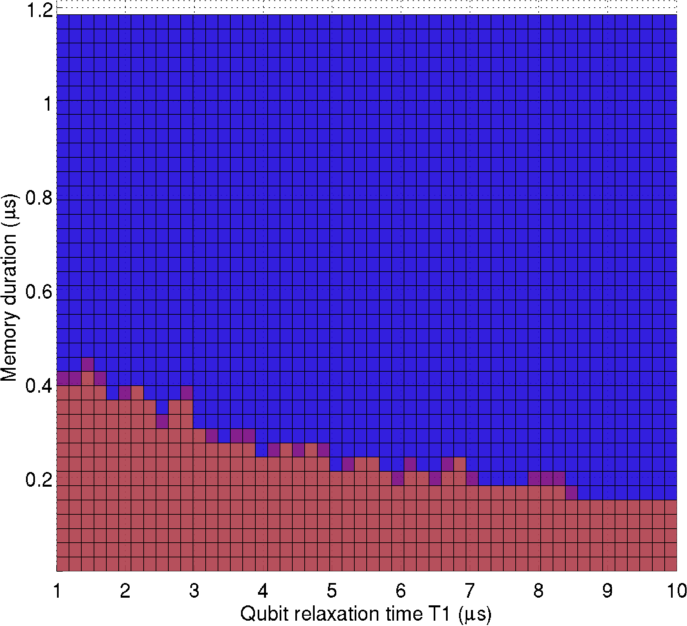} &
\includegraphics[width=0.3\textwidth]{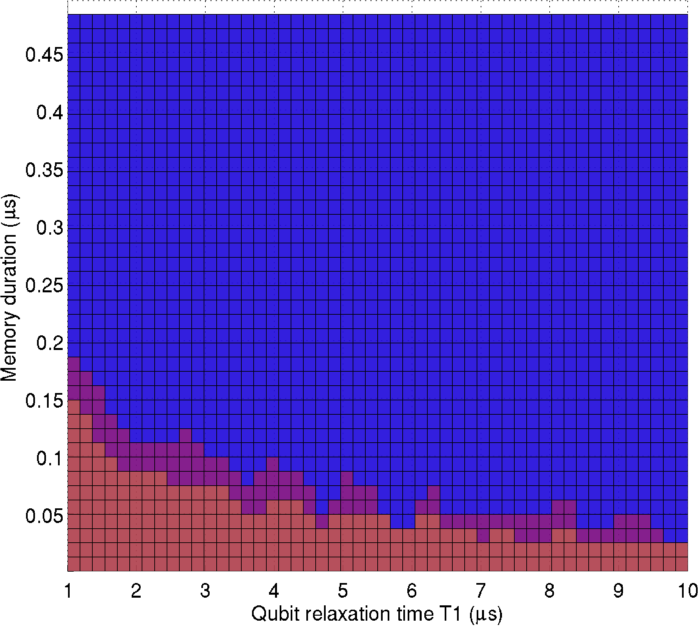}\\
	(c) $SC_D$ & (d) $SC_H$ \\
	\includegraphics[width=0.3\textwidth]{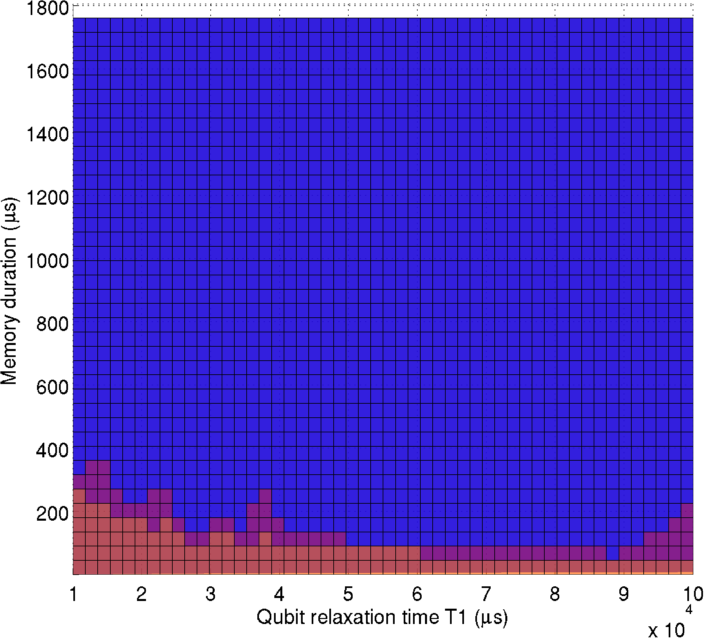}&
	\includegraphics[width=0.3\textwidth]{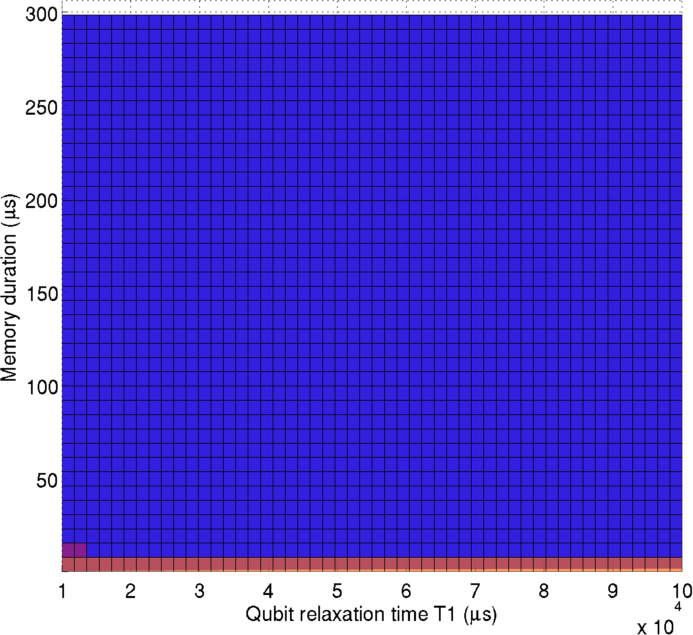} \\
	(e) $IT_S$ & (f) $IT_F$
\end{tabular}
\caption{Plots of $T_1$ time ($\mu$s) versus memory duration ($\mu$s) for six architecture settings under amplitude and phase damping. 
% and damping probability of Surface-17 with real decoherence noise with various architecture settings. 
%Plots represent $+z$-axis view of 3D plots similar to those given in Fig.~\ref{fig_s17_surfaces}. 
The blue and red regions indicate a range of $T_1$ times ($x$-axis) for which encoding a qubit in $\ket{1_L}$ in Surface-17 reduces or increases, respectively, the logical error rate compared to an unencoded $\ket{1}$ qubit in memory for a range durations ($y$-axis). 
% indicates that the amplitude damping probability is higher than the logical error rate, and thus applying Surface-17 results lower error rate than leaving the qubit alone. 
%Note that the memory durations ($x$-axis) and $T_1$ times ($z$-axis) differ significantly for each plot in order to show the crossing line between the amplitude damping probability (unencoded qubit) and the logical error rate (encoded qubit). 
%Also the range of $T_1$ are set differently for each row. 
}\label{fig_s17_architectures}
\end{figure*}

Since the Pauli-twirl approximation aligns well for phase-flip errors, we further compare its performance on bit-flip errors.
Fig.~\ref{fig_s17_surfaces} plots $T_1$ time ($\mu$s) versus memory duration ($\mu$s) versus the logical $X$ failure rate $P_{3,X}$ of a qubit encoded in $\ket{1_L}$ in Surface-17 for the $SC_H$ setting under (a) the Pauli-twirl approximation and (b) amplitude and phase damping.
On the left, the blue surface represents the amplitude damping probability of an unencoded qubit in $\ket{1}$ for a given $T_1$ time and memory duration.
Since the qubit is in $\ket{1}$, phase damping does not apply.
The yellow surface represents the logical error rate $P_{3,X}$ of an encoded qubit for a given $T_1$ time and  surface code round time (see Table \ref{params}).
The orange surface indicates the upper error bar of $P_{3,X}$. 
For the yellow and orange surfaces, the encoded qubit undergoes the surface code three-round window time.  For the blue surface, the unencoded qubit undergoes the given memory duration.

The region where the blue surface lies above the orange and yellow surfaces represents the regime where Surface-17 encoding improves the logical error rate of the qubit (similar to being below pseudothreshold).
The region is larger in Fig.~\ref{fig_s17_surfaces} (b) than Fig.~\ref{fig_s17_surfaces} (a), indicating that Pauli twirling results in a pessimistic estimate of the logical error rate.

The 2D plots on the right are a view from the $+z$-axis. 
The blue and red regions indicate $T_1$ times ($x$-axis) for which encoding a qubit in $\ket{1_L}$ in Surface-17 reduces or increases, respectively, the logical error rate compared to an unencoded $\ket{1}$ qubit in memory for a given duration ($y$-axis). 
%In other words, the blue region represents the regime where a qubit benefits from encoding.
%On the 2D top-down view visually show the region of the relaxation time and the wait time of the qubit where it is expected to have benefits from applying Surface-17. 
The purple region indicates the upper error bar of $P_{3,X}$ where the orange and blue surfaces cross in the 3D plots.
Surface-17 again demonstrates superior performance under amplitude and phase damping compared to Pauli twirling.
For example, for $T_1=1\ \mu$s, memory durations above 150 ns result in lower logical error rates for an encoded qubit than an unencoded qubit, while the Pauli-twirl approximation lowers error rates only for memory durations longer than 350 ns.
%Although the borders on the 2D plots are not very smooth as the data points fluctuates, it is observable that the real Surface-17 performs better under real decoherence than the approximation. 
At $T_1=50\ \mu$s, memory durations of 20 ns result in lower logical error rates, while Pauli twirling indicates lower rates at memory durations longer than 30--70 ns.

\subsection{Amplitude and Phase Damping}
Figure \ref{fig_s17_architectures} shows the same 2D plots for Surface-17 for all six architecture settings under amplitude and phase damping.
%We plot $T_1$ time ($\mu$s) versus the logical $X$ error rate $P_{3,X}$, where the blue and red regions indicate a range of $T_1$ times ($x$-axis) for which Surface-17 reduces or increases, respectively, the logical error rate compared to an unencoded qubit for a range of memory durations ($y$-axis). 
% and damping probability of Surface-17 with real decoherence noise with various architecture settings. 
%The plots represent the $+z$-axis view of 3D plots (not shown here) similar to those given in Fig.~\ref{fig_s17_surfaces}. 
% indicates that the amplitude damping probability is higher than the logical error rate, and thus applying Surface-17 results lower error rate than leaving the qubit alone. 
%The memory durations ($x$-axis) and $T_1$ times ($z$-axis) differ significantly for each plot in order to show the crossing line between the amplitude damping probability (unencoded qubit) and the logical error rate (encoded qubit)
%
For each architecture, the $y$-axis ranges from 0 $\mu$s to the time per surface code window.
%The dependence of error correction performance on the per-round and $T_1$ times is apparent.
In all graphs, we see that as $T_1$ increases, encoding improves the logical error rate for a larger range of memory durations.
This behavior is expected since the amplitude damping probability monotonically increases with memory duration.

In Fig.~\ref{fig_s17_architectures}(a), at $T_1=1\ \mu$s we observe that for the $SC_S$ parameters, encoding does not improve the logical error rate for any plotted memory duration.
However, with 10 times faster gates ($SC_F$), we see performance improvement, as shown in Fig.~\ref{fig_s17_architectures}(b).  
At $T_1=1\ \mu$s, encoding provides a better logical error rate than an unencoded qubit in memory for at least $8\ \mu$s.
At $T_1=10\ \mu$s, $SC_S$ shows no improvement with encoding, while $SC_F$ exhibits improvements for memories of $2\ \mu$s or longer.

The $SC_H$ setting accounts for 100 times faster preparation and measurement than $SC_F$ and roughly 10 times faster gates. 
The faster times lead to significantly better performance under encoding. 
For example, at $T_1=1\ \mu$s and $10\ \mu$s in Fig.~\ref{fig_s17_architectures}(d), the logical error rate decreases due to encoding for memory durations longer than $150\ $ns and $20\ $ns, respectively.

Comparing Fig.~\ref{fig_s17_architectures}(c) and (d), we find CNOT time strongly influences performance.
A CNOT gate is four times longer in $SC_D$ than $SC_H$.
The longer two-qubit gate time is reflected in the poorer performance of Surface-17 under $SC_D$ parameters.
At $T_1=1\ \mu$s, $SC_D$ only indicates logical error rate reduction due to encoding at memory durations roughly 3 times longer than those required for $SC_H$.
%to roughly indicates improvements from encoding at memory durations longer than 0.4 $\mu$s, .

Fig.~\ref{fig_s17_architectures}(e) and (f) show similar results for the ion trap settings.
While $IT_F$ assumes 10 times faster CNOT gates, both $IT_S$ and $IT_F$ yield lower logical error rate upon encoding for a range of $T_1$ times.
$IT_S$ results in improvements for memory durations longer than 300--400 $\mu$s, while $IT_F$ results in improvements for memory durations above around 15 $\mu$s.

\begin{figure*}[htb!] %\centering
\begin{tabular}{ c c c }

	\includegraphics[width=0.3\textwidth]{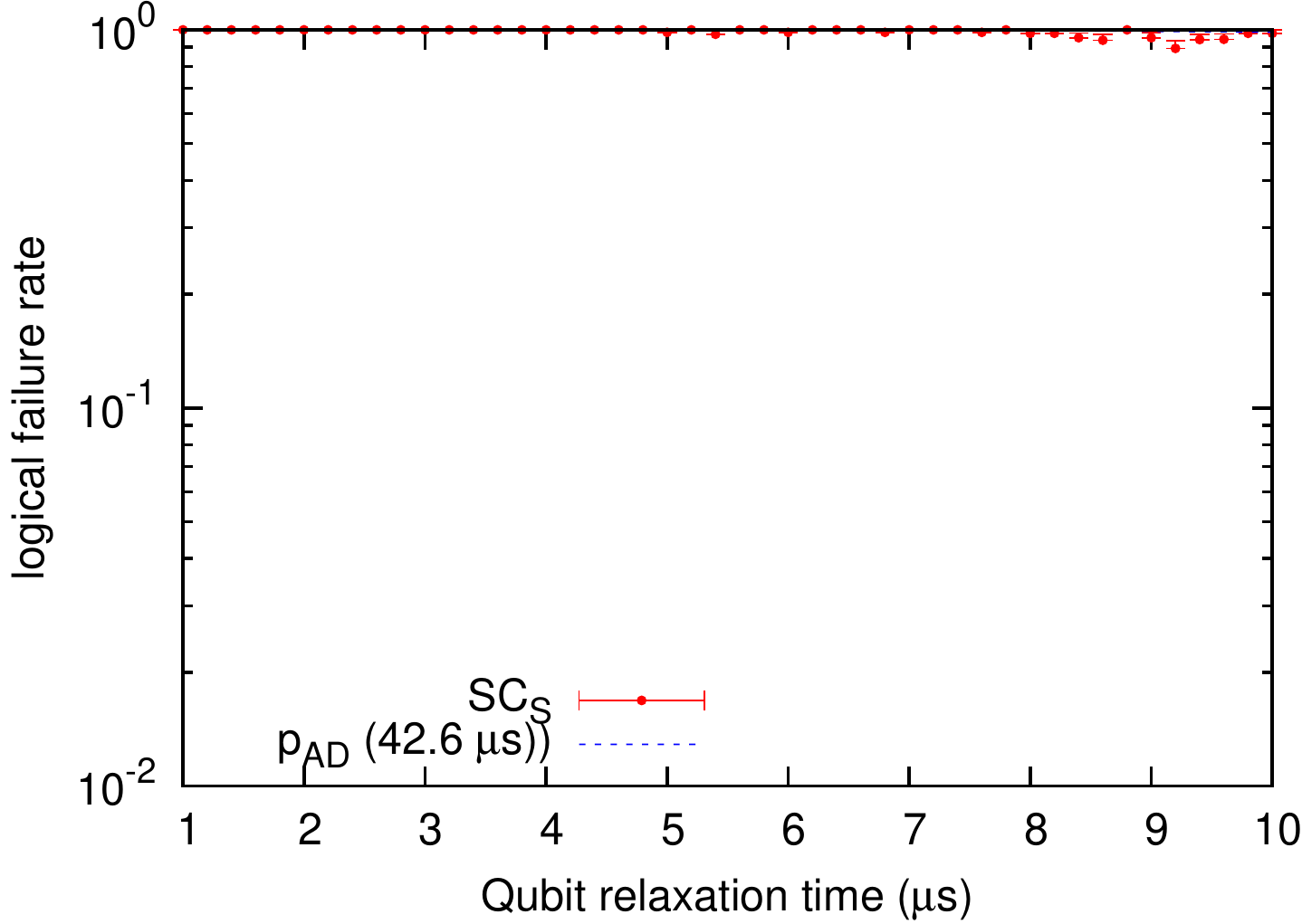}&
	\includegraphics[width=0.3\textwidth]{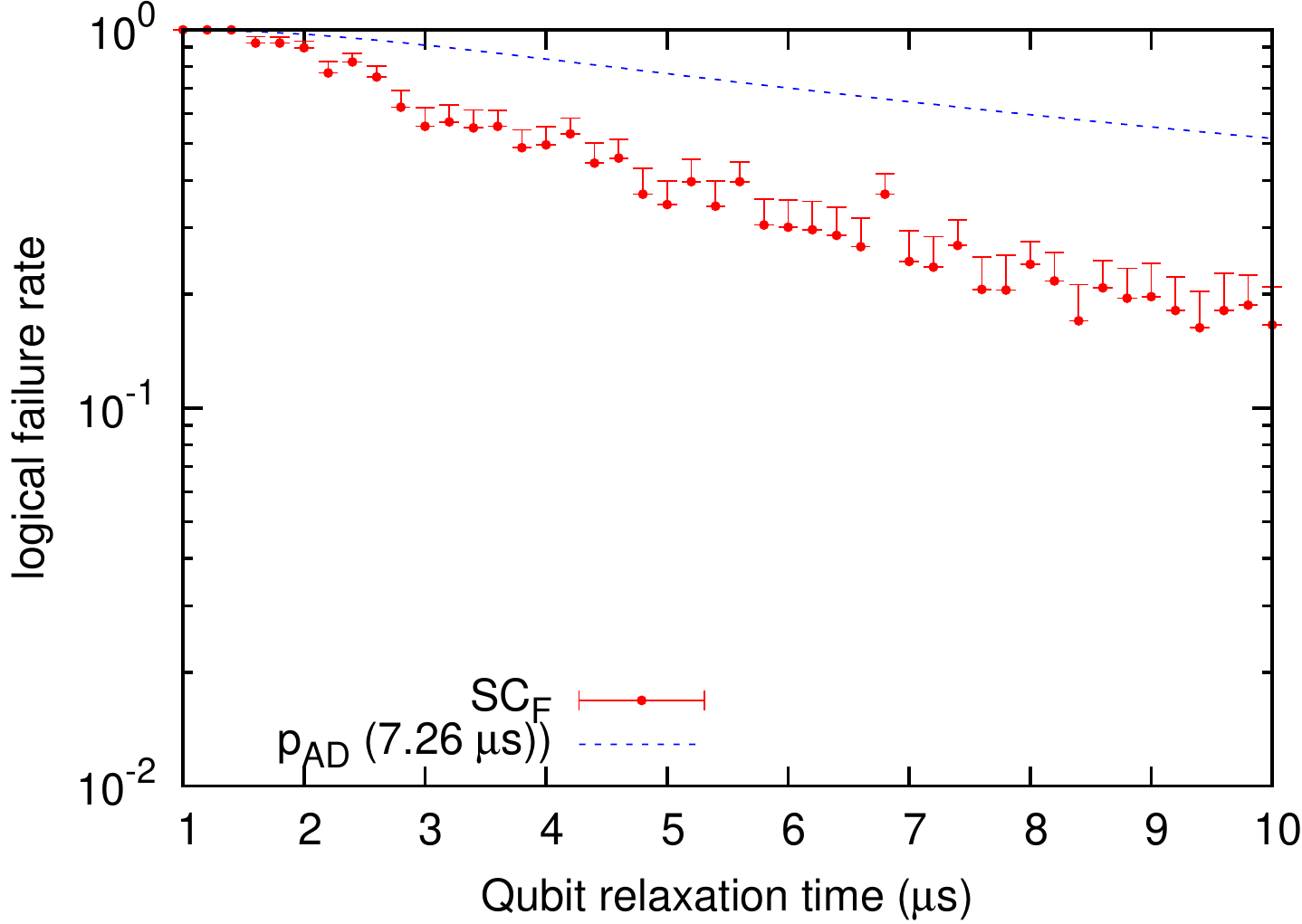} \\	

	(a) $SC_S$ & (b) $SC_F$ \\
	\includegraphics[width=0.3\textwidth]{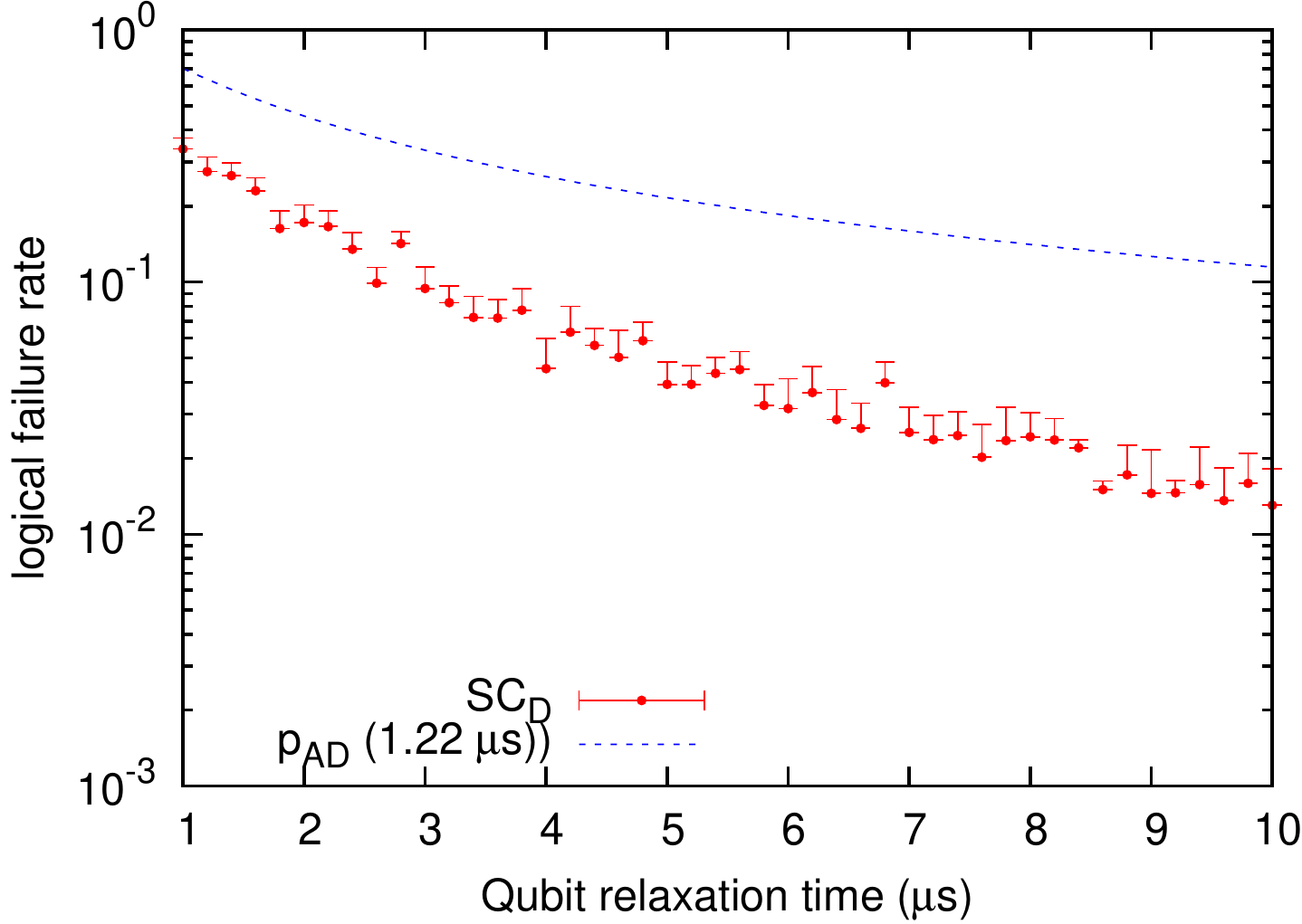} &
\includegraphics[width=0.3\textwidth]{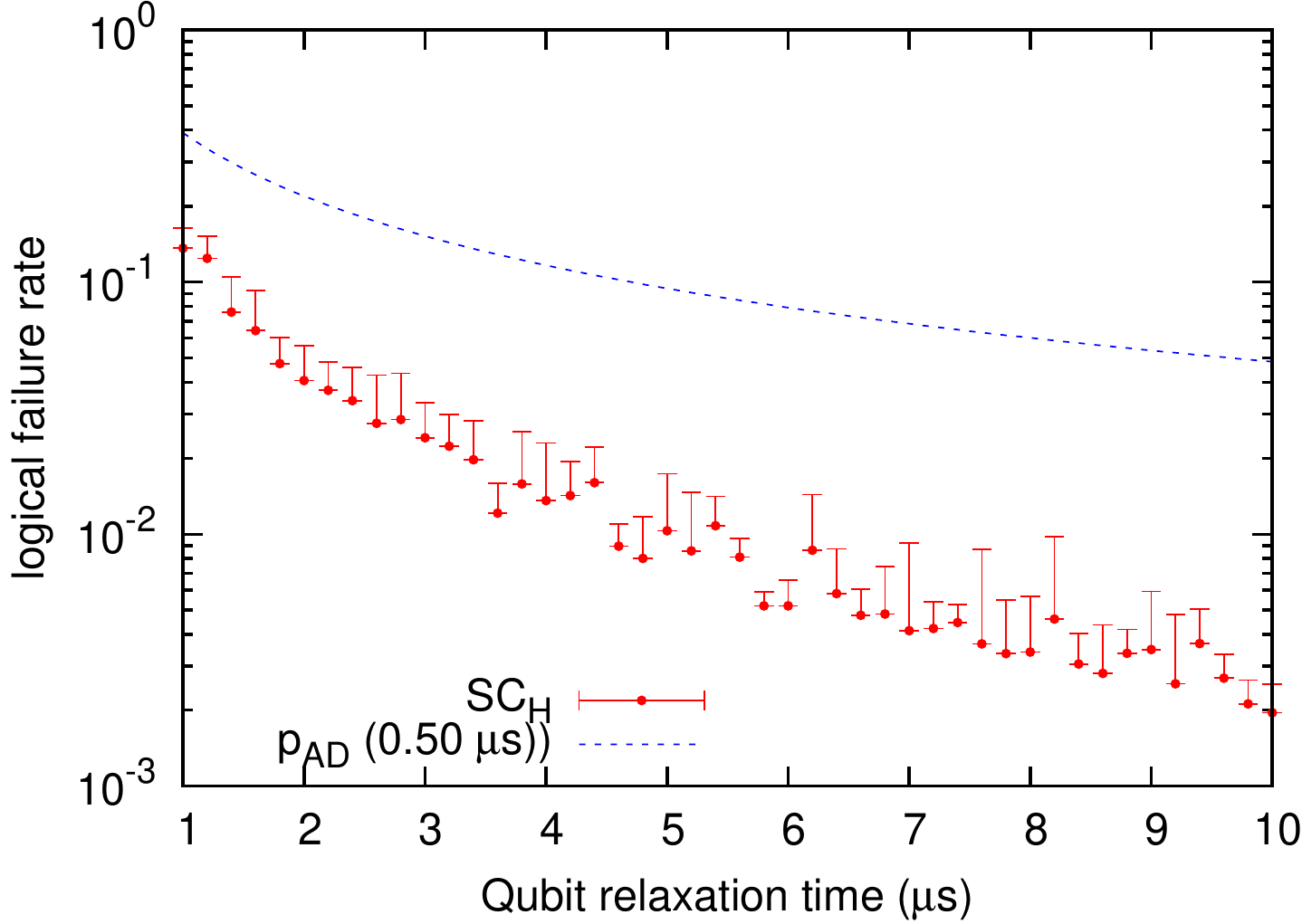}\\
	(c) $SC_D$ & (d) $SC_H$ \\
	\includegraphics[width=0.3\textwidth]{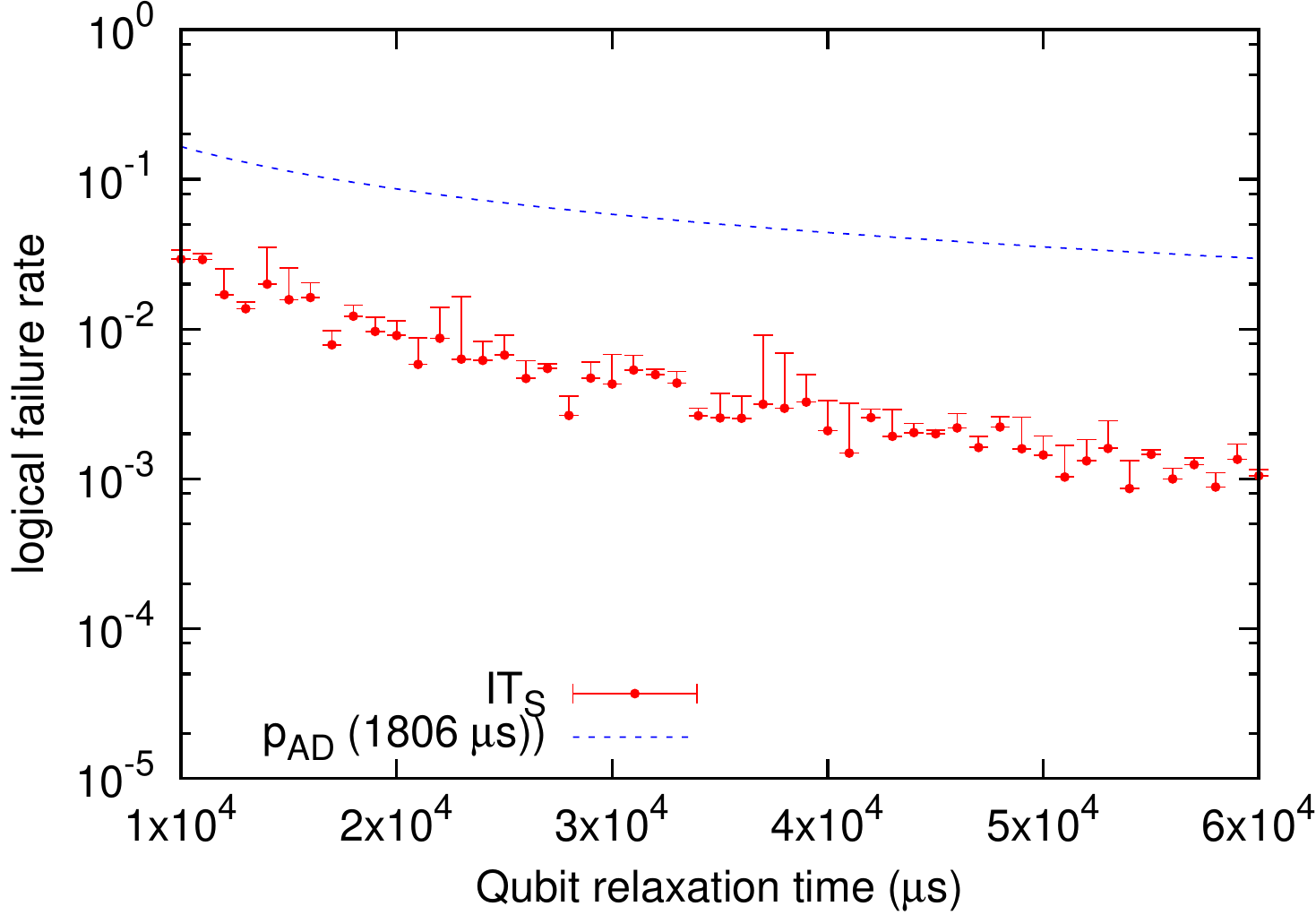}&
	\includegraphics[width=0.3\textwidth]{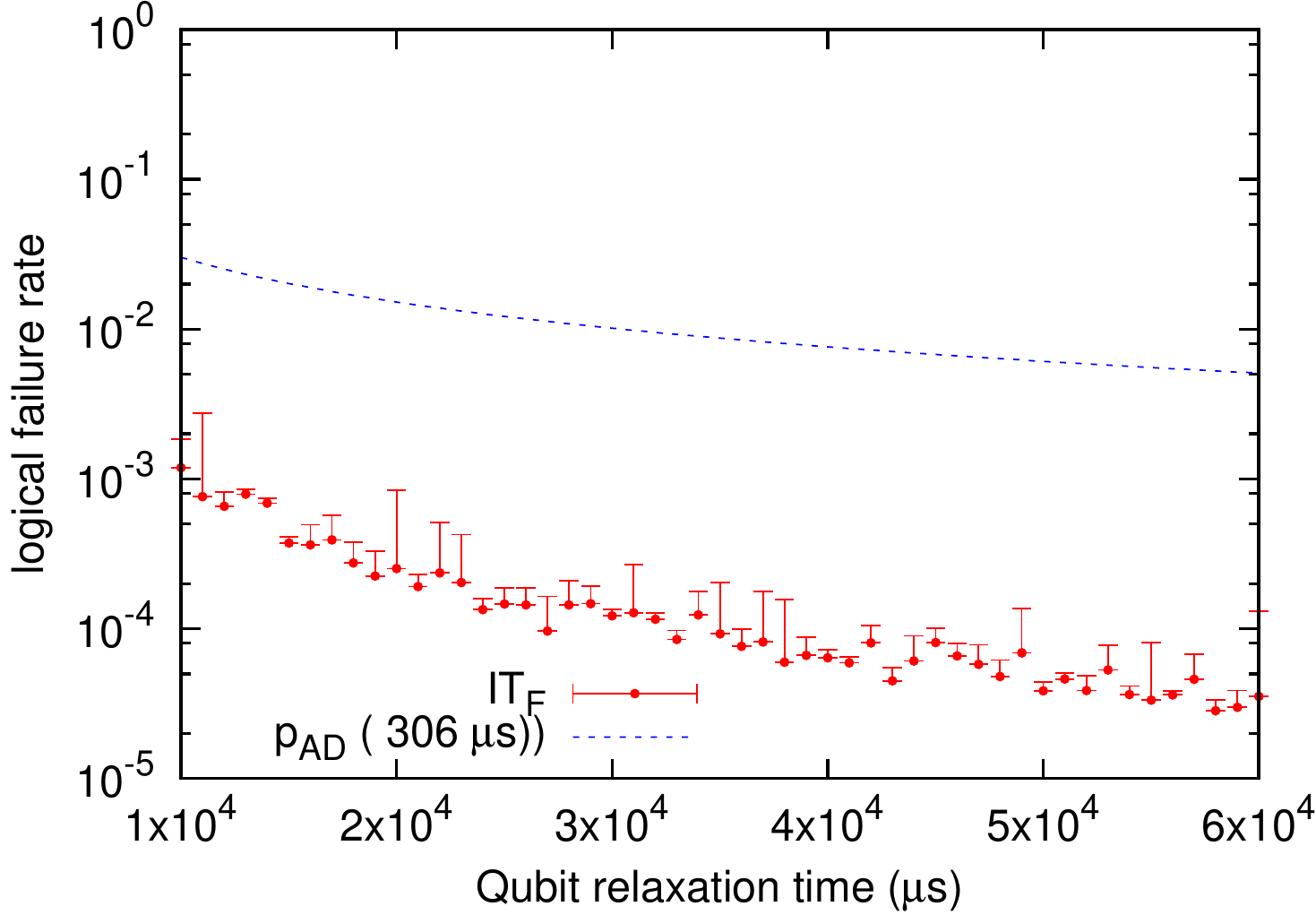} \\
	(e) $IT_S$ & (f) $IT_F$
\end{tabular}
\caption{Plots of $T_1$ time ($\mu$s) versus the logical $X$ error rate $P_{3,X}$ for six architecture settings  for a qubit encoded in $\ket{1_L}$ in Surface-17 subject to amplitude and phase damping (red) and an unencoded $\ket{1}$ qubit subject to amplitude damping (blue) for the duration of three rounds of the surface code.
% and damping probability of Surface-17 with real decoherence noise with various architecture settings. 
%Plots represent $+z$-axis view of 3D plots similar to those given in Fig.~\ref{fig_s17_surfaces}. 
%The blue and red regions indicate a range of $T_1$ times ($x$-axis) for which Surface-17 reduces or increases, respectively, the logical error rate compared to an unencoded qubit for a range of memory durations ($y$-axis). 
% indicates that the amplitude damping probability is higher than the logical error rate, and thus applying Surface-17 results lower error rate than leaving the qubit alone. 
%Note that the memory durations ($x$-axis) and $T_1$ times ($z$-axis) differ significantly for each plot in order to show the crossing line between the amplitude damping probability (unencoded qubit) and the logical error rate (encoded qubit). 
%Also the range of $T_1$ are set differently for each row. 
}\label{fig_s17_errors}
\end{figure*}

In Figure \ref{fig_s17_errors}, we plot qubit relaxation time $T_1$ ($\mu$s) versus logical error rate $P_{3,X}$ (red) or amplitude damping probability (blue) for the six architecture settings (analogous to Fig.~\ref{fig_s17_real_vs_pauli}(b)).
All plots assume a three-round memory duration.
From the plots, the logical error rate for each architecture for a given $T_1$ time can be extracted.
As gate times and $T_1$ times improve, the logical error rate decreases.  An order of magnitude improvement in logical error rate can be obtained, for example, in improving gates time from those of $SC_F$ to those of $SC_H$.

The plots also indicate that near-term experiments may be able to detect improved logical error rates due to encoding, providing experimental evidence of surface code error correction.
For example, for $T_1=1$ $\mu$s, no difference in the logical error rate can be detected between an encoded and unencoded qubit given settings $SC_S$ and $SC_F$.  
However, $SC_F$ exhibits differences on the order of one magnitude at $T_1$ times larger than $30\ \mu$s.
Both $SC_D$ and $SC_H$ settings indicate significant difference in the logical error rate on an encoded versus unencoded qubit.
In the case of both ion trap settings, a difference in logical error rate can be detected starting at $T_1=10$ $\mu$s.
%  for $SC_H$ and $IT_F$, error rates at short $T_1$ times are significantly different between an encoded and unencoded qubit.

%The plots in Fig.~\ref{fig_s17_errors} also give the logical error rate values for .
%We see the logical error rate decreases

%These plots show the  the difference between the blue surface and 

\begin{figure*}[tb!] %\centering
\begin{tabular}{ c c c }

	\includegraphics[width=0.3\textwidth]{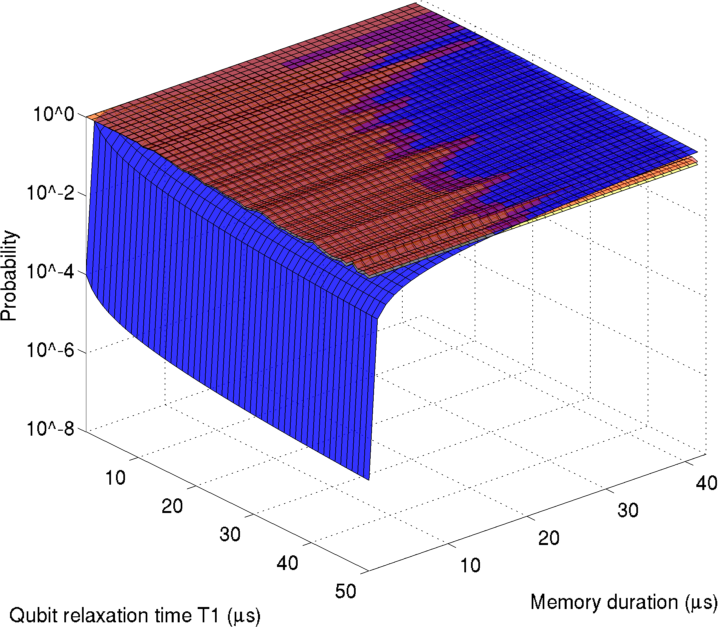}&
	\includegraphics[width=0.3\textwidth]{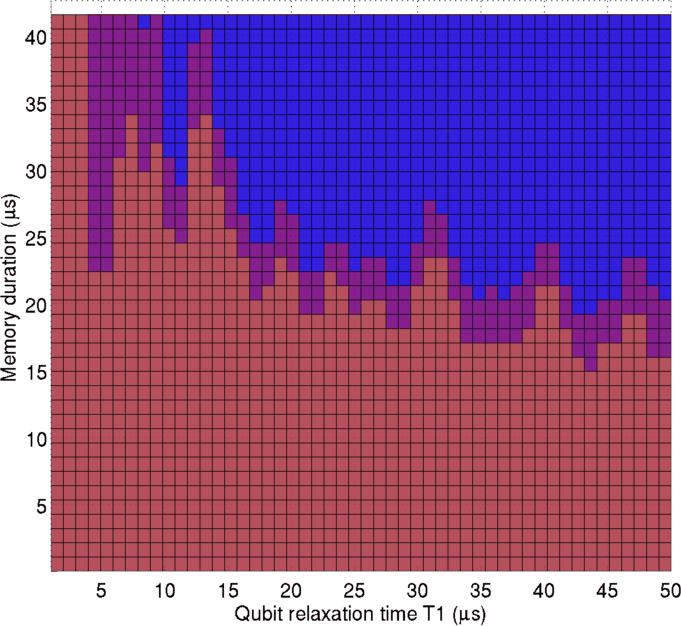}&
     	\includegraphics[width=0.3\textwidth]{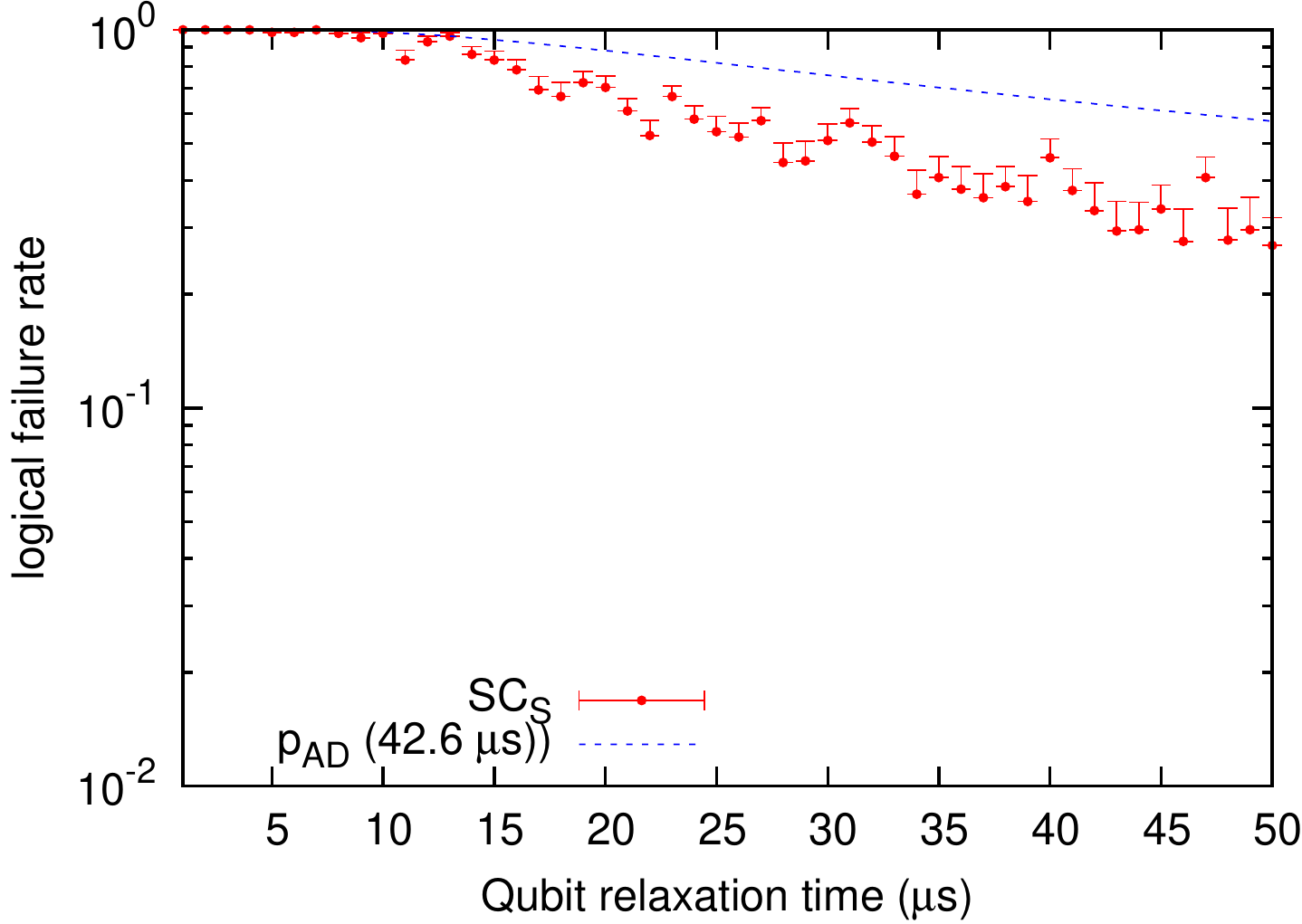}\\
	& (a) $SC_S$ & \\

	\includegraphics[width=0.3\textwidth]{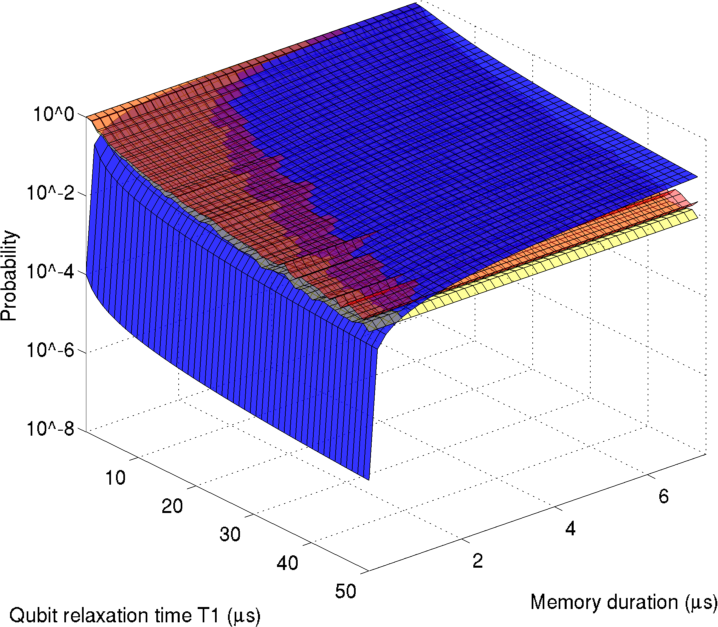} &
	\includegraphics[width=0.3\textwidth]{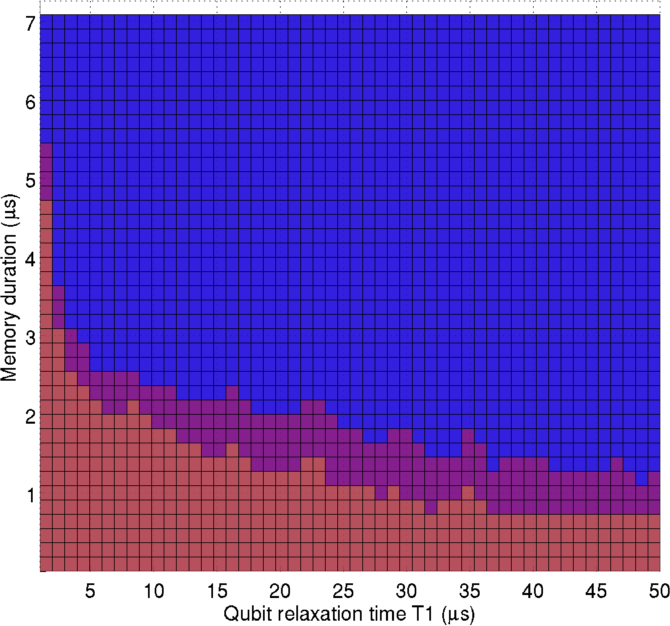} &
	\includegraphics[width=0.3\textwidth]{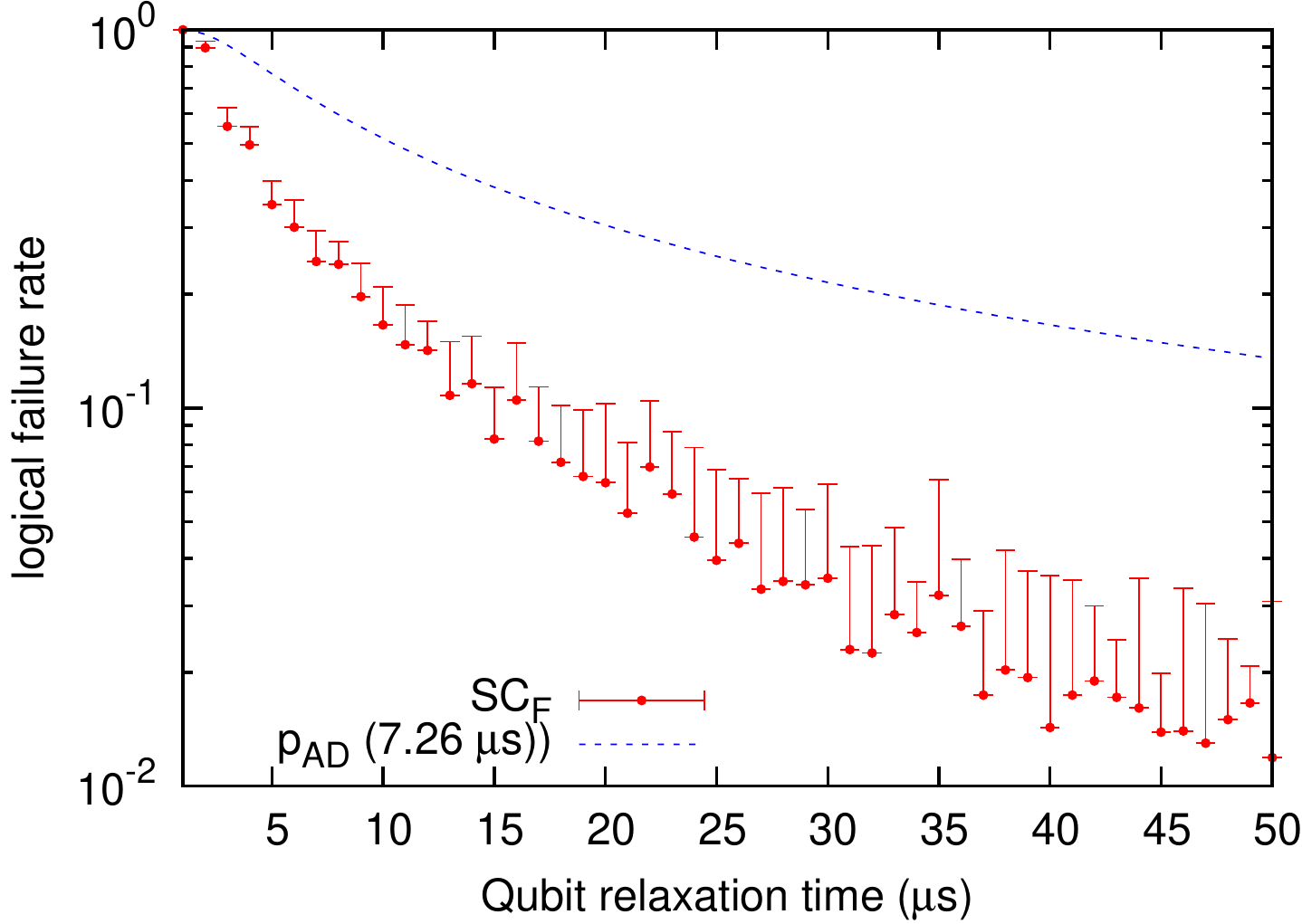} \\
	& (b) $SC_F$ & \\

	\includegraphics[width=0.3\textwidth]{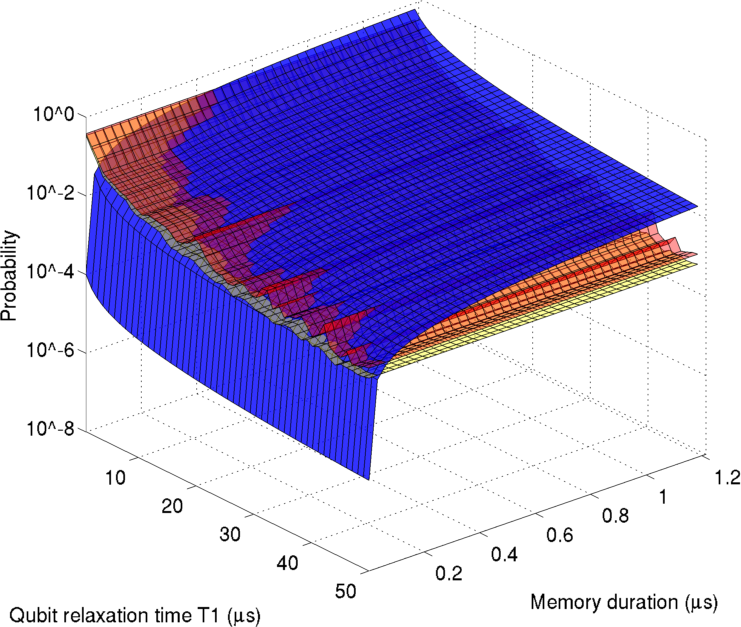} &
	\includegraphics[width=0.3\textwidth]{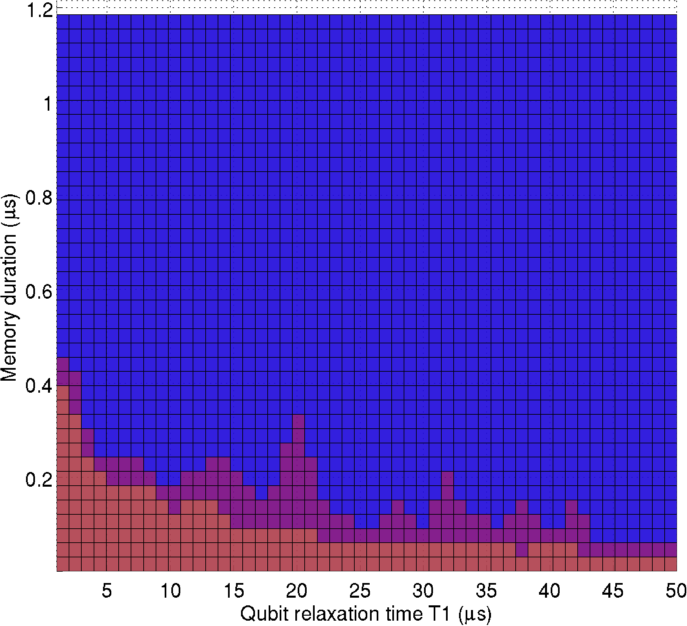} &
	\includegraphics[width=0.3\textwidth]{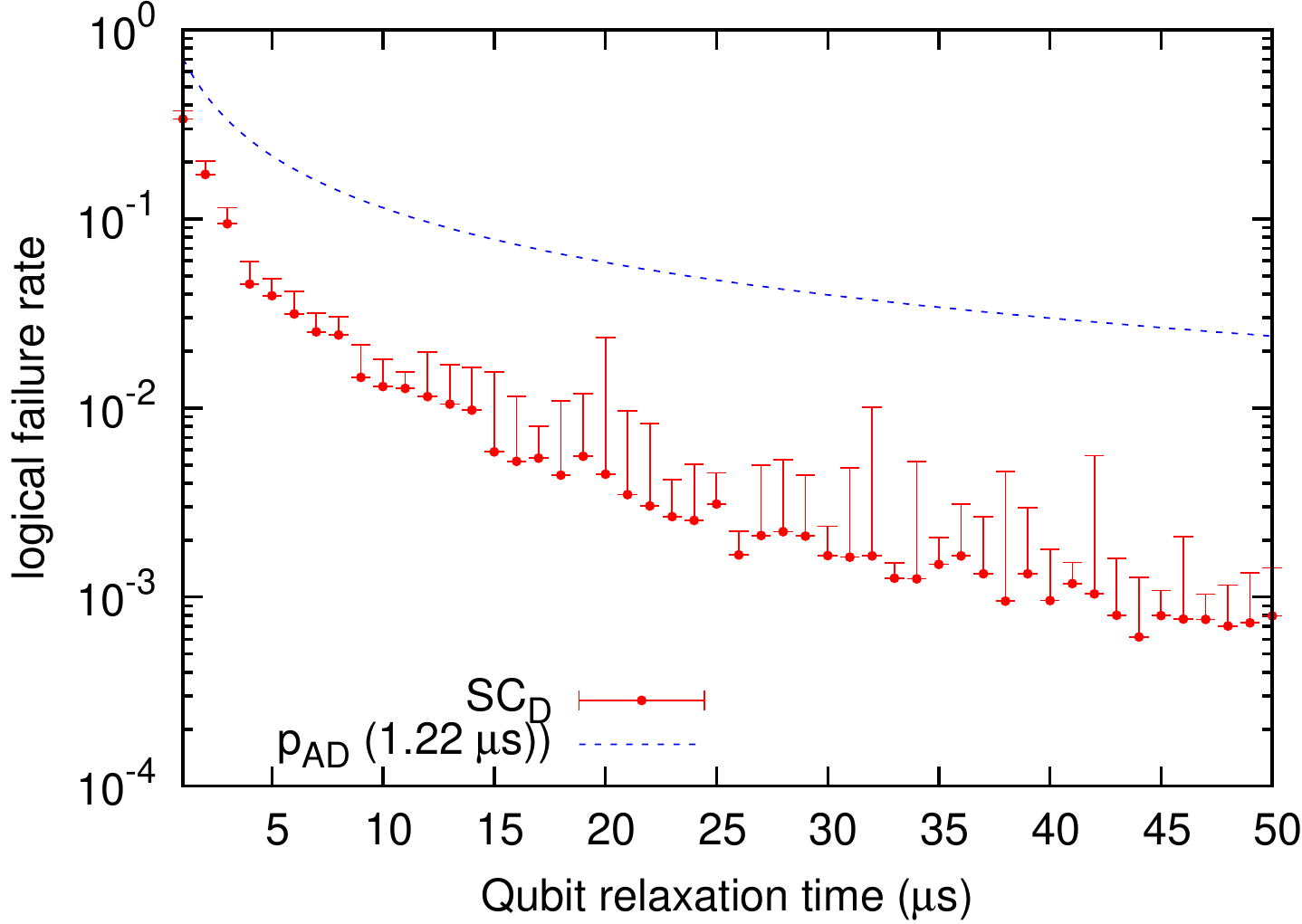} \\
	& (c) $SC_D$ & \\

	\includegraphics[width=0.3\textwidth]{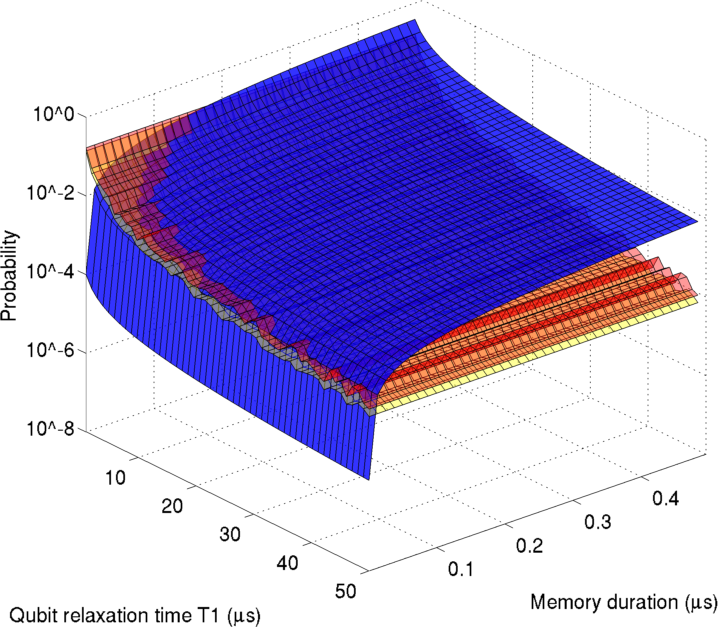}&
	\includegraphics[width=0.3\textwidth]{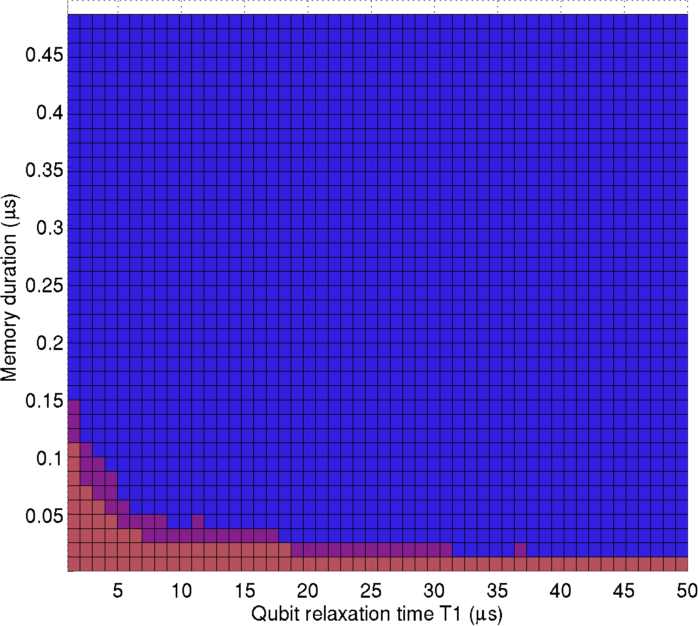}&
	\includegraphics[width=0.3\textwidth]{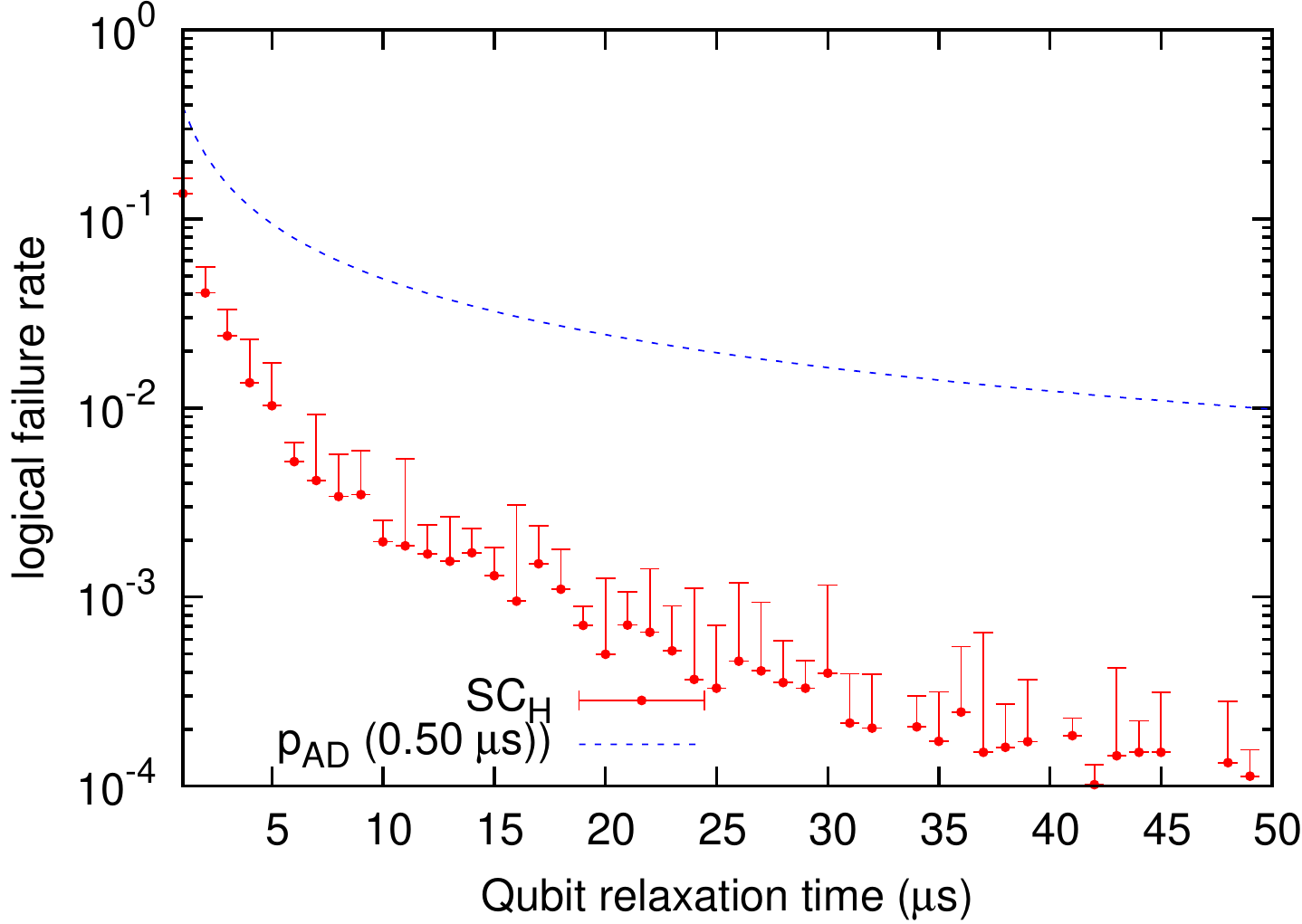}\\
	& (d) $SC_H$ &
\end{tabular}
\caption{Plots for four superconductor architecture settings under amplitude and phase damping for $T_1$ times up to $50\ \mu$s. 
(Left) Plots of $T_1$ time ($\mu$s) versus memory duration ($\mu$s) versus logical failure rate. 
(Middle) Plots of $T_1$ time ($\mu$s) versus memory duration ($\mu$s).
(Right) Plots of $T_1$ time ($\mu$s) versus logical failure rate.
The blue and red regions indicate a range of $T_1$ times ($x$-axis) for which encoding a qubit in $\ket{1_L}$ in Surface-17 reduces or increases, respectively, the logical error rate compared to an unencoded $\ket{1}$ qubit in memory for a range durations ($y$-axis). 
% indicates that the amplitude damping probability is higher than the logical error rate, and thus applying Surface-17 results lower error rate than leaving the qubit alone. 
%Note that the memory durations ($x$-axis) and $T_1$ times ($z$-axis) differ significantly for each plot in order to show the crossing line between the amplitude damping probability (unencoded qubit) and the logical error rate (encoded qubit). 
%Also the range of $T_1$ are set differently for each row. 
}\label{fig_s17_architectures_50us}
\end{figure*}

Figure \ref{fig_s17_architectures_50us} contains plots for $T_1$ times up to $50\ \mu$s for the four superconducting architectures.
The left column contains 3D plots of $T_1$ time ($\mu$s) versus memory duration ($\mu$s) versus logical failure rate.
The middle column contains 2D plots of $T_1$ time ($\mu$s) versus memory duration ($\mu$s).
The right column contains 2D plots of $T_1$ time ($\mu$s) versus logical failure rate.

We conclude that gate durations in the $SC_S$ setting are too slow for Surface-17 to significantly decrease the logical error rate given realistic $T_1$ times (Fig.~\ref{fig_s17_architectures_50us}(a)).
However, given gate durations between the $SC_F$ and $SC_H$ settings and current $T_1$ times of $20$--$40\ \mu$s, encoding a qubit in Surface-17 results in significantly improved error rates over an unencoded qubit (Fig.~\ref{fig_s17_architectures_50us}(b)--(d)).
For both superconductor and ion trap architectures, near-term experimental implementations could demonstrate surface code error correction of a single logical qubit, and measure signficiant improvements in the logical error rate.
We find that previous estimates of 2.6--2.8 $\mu$s $T_1$ times \cite{Ghosh2012} to achieve improved logical error rates are too high, and in fact at only 1 $\mu$s $T_1$ time, the logical error rate can be improved using Surface-17. 

\section{Conclusion}\label{sec:conclude}

We have analyzed three distance-three surface code layouts under realistic noise models.
Under symmetric depolarizing noise, we find the pseudothreshold is slightly lower for Surface-13 as compared to Surface-17 and 25. 
%Due to its slightly better performance and use of 30\% fewer resources, we determine Surface-17 is the preferred layout.
%
We have compared the performance of Surface-17 simulated under a Pauli-twirl approximation and amplitude and phase damping.
%With the \Liquid software architecture we directly simulated the Surface-17 circuit under the latter (non-Clifford) channel.
Our results show that Pauli twirling pessimistically estimates the logical bit-flip rate.
%The approximation worsens as the qubit relaxation time $T_1$ increases. 
Thus the surface code threshold under realistic noise may be significantly better than previously calculated.
%, especially for architectures with longer relaxation times such as ion traps.

We have also simulated the 17-qubit surface code under amplitude and phase damping for six architecture settings. 
%We have compared the logical error rate of surface code with amplitude damping probability of a qubit using 3D surface plots. 
%The intersection between those two surfaces indicate the region of qubit waiting time and the qubits' relaxation time where application of the surface code does not harm or benefit. 
%The blue region with longer relaxation time and longer duration from the intersection suggest that the surface code has higher chance to reduce the total error rate. 
While gate durations in the $SC_S$ setting are too slow, gate durations between $SC_F$ and $SC_H$ with current $T_1$ times show improved logical error rates for a qubit encoded in Surface-17.
For both superconductor and ion trap architectures, current state-of-the-art experiments may be able to demonstrate surface code error correction.
%For example, with $T_1=1$ $\mu$s and $SC_H$ settings, logical error rates will improve by encoding in Surface-17 and may be detected in experiment.
For example, with $T_1$ around $10\ \mu$s and $SC_F$ settings, logical error rates will improve by encoding in Surface-17 and may be detected in experiment.
With $SC_H$ settings, even shorter $T_1$ times will result in significant improvements with encoding.

Methods of approximating decoherence using Clifford gates have recently been shown to be more accurate than Pauli twirling \cite{Mauricio2013,Cory012013}. 
However, studies have only been conducted at the gate operation level as opposed to the circuit level of a given code.
A direction for future work is to simulate these noise models on Surface-17 to compare to amplitude and phase damping.
Another direction is to determine the performance of Surface-17 under leakage.
%Another direction is to determine the computation threshold of the surface code under realistic noise models, in particular for the Hadamard and CNOT gates.
Finally, development and simulation of realistic noise models for specific architectures will be important for guiding experimental surface code implementations.
%KMS:WHAT VALUES SHOULD EXPERIMENTALISTS TARGET TO ENSURE CORRECTION AND A CLEANER OUTGOING QUBIT FOR THESE LAYOUTS? YT: added one sentence at the end of Results section. Tried a simple table can this kind of table help? 

\begin{acknowledgments}
We are especially grateful to Dave Wecker for helping with the surface code and noise model implementations in \LiquidB.  We thank Ken Brown, Leo DiCarlo, and Andrew Landahl for valuable discussions.  We also thank Leo Kouwenhoven, Lieven Vandersypen, Austin Fowler, and Blake Johnson for useful feedback.
\end{acknowledgments}

\bibliography{surface_paper_ref}% Produces the bibliography via BibTeX.

%\newpage %Just because of unusual number of tables stacked at end

%\appendix
%\section{Logical Error Rates for Architecture Settings}

\end{document}